\documentclass{aa}  

\usepackage{graphicx}
\usepackage{txfonts}
\usepackage{lscape}
\begin{document} 

   \title{Shapes of stellar activity cycles}

   \author{T. Willamo
          \inst{1}
          \and
          T. Hackman\inst{1}
          \and
          J. J. Lehtinen\inst{2}
          \and
          M. J. Käpylä\inst{3}$^,$ \inst{2}
          \and
          N. Olspert\inst{2}
          \and
          M. Viviani\inst{2}
          \and
          J. Warnecke\inst{2}
          }

   \institute{Department of Physics, P.O. Box 64, FI-00014 University of Helsinki, Finland \\
              \email{teemu.willamo@helsinki.fi}
             \and
             Max Planck Institute for Solar System Research, Justus-von-Liebig-Weg 3, D-37077 Göttingen, Germany
             \and
             Department of Computer Science, Aalto University, PO Box 15400, FI-00076 Aalto, Finland
             }

   \date{Received February 5, 2020; accepted April 17, 2020}

 
  \abstract
   {Magnetic activity cycles are an important phenomenon both in the Sun and other stars. The shape of the solar cycle is commonly characterised by a fast rise and a slower decline, but not much attention has been paid to the shape of cycles in other stars.}
   {Our aim is to study whether the asymmetric shape of the solar cycle is common in other stars as well, and compare the cycle asymmetry to other stellar parameters. We also study the differences in the shape of the solar cycle, depending on the activity indicator that is used. The observations are also compared to simulated activity cycles.}
   {We used the chromospheric \ion{Ca}{ii} H\&K data from the Mount Wilson Observatory HK Project. In this data set, we identified 47 individual cycles from 18 stars. We used the statistical skewness of a cycle as a measure of its asymmetry, and compared this to other stellar parameters. A similar analysis has been performed for magnetic cycles extracted from direct numerical magnetohydrodynamic simulations of solar-type convection zones.}
   {The shape of the solar cycle (fast rise and slower decline) is common in other stars as well, although the Sun seems to have particularly asymmetric cycles. Cycle-to-cycle variations are large, but the average shape of a cycle is still fairly well represented by a sinusoid, although this does not take its asymmetry into account. We find only slight correlations between the cycle asymmetry and other stellar parameters. There are large differences in the shape of the solar cycle, depending on the activity indicator that is used. The simulated cycles differ in the symmetry of global simulations that cover the full longitudinal range and are therefore capable of exciting non-axisymmetric large-scale dynamo modes, and wedge simulations that cover a partial extent in longitude, where only axisymmetric large-scale modes are possible. The former preferentially produce positive and the latter negative skewness.}
   {}


   \maketitle
%

\section{Introduction}

The shape of the 11-year sunspot cycle is not perfectly symmetric, but characterised by a faster rise from minimum to maximum and a slower decline from maximum to minimum \citep{waldmeier1935}. Another common feature that deviates from a sinusoid shaped cycle is the typical double peak, known as the Gnevyshev gap \citep{gnevyshev1963}. \citet{gnevyshev1967,gnevyshev1977} suggested that the solar cycle generally consists of two waves of activity. There is an asymmetry in solar activity between the northern and southern hemisphere \citep[e.g.][]{newton1955,deng2016}. 
\citet{norton2010} studied the solar cycle separately on each hemisphere and concluded that differences in the hemispheres cannot explain the Gnevyshev gap, but a mechanism must be producing it for both hemispheres. One factor that might affect it is the complexity of active regions. Simple active regions, with unipolar or bipolar sunspot groups, on average appear earlier in the solar cycle than more complex active regions \citep{nikbakhsh2019}. Thus the simple regions dominate the first peak of the maximum, and the complex regions only have a notable effect on the latter peak. \citet{feminella1997} found that the activity dip in the Gnevyshev gap is more evident in high-energy phenomena, such as the occurrence of long-lasting energetic flares, while the occurrence of flares and other phenomena with lower energies tend to follow the simple 11-year cycle. 
The cycle amplitude and length of the rising phase are also anti-correlated. This is known as the Waldmeier effect \citep{waldmeier1935,waldmeier1939}.

There is no reason for stellar analogues of the solar cycle to be perfectly symmetric either, but they are usually fitted with simple sinusoids, and not much attention has been paid to their shape. \citet{reinhold17} showed that cycles derived from the variability of Kepler stars deviate from simple sinusoids, the average shape showing a sharp maximum and flattened minimum. The authors also hypothesised that this effect might have a temperature dependence because it was weak for the coolest stars.

The solar cycle has been modelled with many different mathematical formulations accounting for their asymmetry \citep{nordemann1992,elling1992,hathaway1994,volobuev2009,du2011}. \citet{takalo2018} applied the principal component analysis to the solar cycle and divided it into two components, an average cycle component, which always has the same shape, with varying period and amplitude, and one component that varies from cycle to cycle.

One parameter that has been used to measure asymmetries of solar cycles is the skewness. This is a measure of asymmetry that is commonly used in statistics. 
\citet{ramaswamy1977} reported a relation between the ratio of the maximum sunspot number of the following cycle to the current cycle $\mu$ and the skewness $\gamma$ of the current cycle as

\begin{equation}
\gamma + 0.37\mu = 0.80.
\end{equation}

\noindent \citet{lantos2006} improved the correlation by separately considering even and odd cycles, and derived the following formulae:

\begin{equation}
\mu \!=\!
\left\{
{\begin{array}{rl}
-2.1092\gamma + 1.9418\,& \textrm{when the current cycle is even},\\
-1.2552\gamma + 1.3570\,& \textrm{when the current cycle is odd}.
\end{array}}
\right.
\end{equation}

\noindent Stellar cycles, however, have not been modelled as extensively. \citet{garg2019} found the Waldmeier effect in stars from Mount Wilson observatory data. They also studied stellar cycle asymmetries by fitting similar functions as for the solar cycle, and they calculated the skewness. \citet{pipin2016} found from numerical mean-field simulations for solar-type stars that magnetic cycles of a higher amplitude are more asymmetric, until at some amplitude, the asymmetry becomes saturated. 

The methods that are commonly used to study stellar cycles are not capable of taking cycle asymmetries into account because usually the cycles are assumed to have sinusoidal form. 
The Lomb-Scargle periodogram \citep{lomb1976,scargle82} is a commonly used method, but it assumes a strict periodicity, which is generally not the case in stellar activity cycles. The duration of the solar cycle, for instance, varies from about 8 years to 14 years. The use of quasi-periodic models allows the cycles not to be strictly periodic \citep{olspert2018}. Here, we study each cycle individually to account for cycle-to-cycle differences in the duration and shape of the cycles.

\section{Data}

\subsection{Mount Wilson data}

We used the publicly available \ion{Ca}{ii} H\&K S-index measurements from the Mount Wilson (MW) Observatory, a programme started by \citet{wilson1978}. The data set, including almost 2300 stars, was gathered between 1966 and 1995, with additional data for 35 stars extended to 2001. The S-index, defined as

\begin{equation}
  S = \alpha \frac{H+K}{V+R},
\end{equation}

\noindent is a sensitive indicator of chromospheric magnetic activity \citep[e.g.][]{egeland2017}. Here $H$ and $K$ indicate flux integrated over narrow passbands centred around the \ion{Ca}{ii} H and K line cores, and $V$ and $R$ are broad continuum bands on the violet and red sides of the Ca lines. $\alpha$ is a calibration factor that is determined for each night from standard lamp and standard star observations.

\citet{baliunas1995} determined the periodicity of the MW stars with Lomb-Scargle periodograms, and divided the stars with cyclic variations into four different categories based on the false-alarm probability (FAP), the probability that a peak as strong as the observed peak would randomly occur in the Lomb-Scargle periodogram, assuming purely Gaussian noise. These categories are labelled \textquoteleft excellent\textquoteright, \textquoteleft good\textquoteright, \textquoteleft fair', and \textquoteleft poor\textquoteright, corresponding to $FAP \leq 10^{-9}$, $10^{-9} < FAP \leq 10^{-5}$, $10^{-5} < FAP \leq 10^{-2}$, and $10^{-2} < FAP \leq 10^{-1}$, expressed in percent, respectively. The authors note, however, that because of variations due to the growth and decay of active regions, for instance, which is non-Gaussian noise, the FAP should not be taken too literally.

\citet{olspert2018} compared the cycle periods in \cite{baliunas1995} and periods derived with quasi-periodic methods. They found that the results were similar for the \textquoteleft excellent' stars, while the resemblance weakens gradually for the \textquoteleft good', \textquoteleft fair', and \textquoteleft poor' stars. 
Some of the differences, however, can be explained by their use of additional data from the extended 2001 data set, and by the higher significance level.

In our sample we included all the stars defined as \textquoteleft excellent' or \textquoteleft good' by \citet{baliunas1995}, with the exception of HD 78366,  HD 201092, and HD 156206, which are labelled \textquoteleft good'. HD 78366 is left out because it is not clear where its minima are because there are multiple secondary minima in the data. HD 201092 was excluded because its minimum around JD-2444000=2500 is very difficult to define; there seems to be a local maximum where the minimum should be according to the 11.7-year cycle reported by \citet{baliunas1995}. They found no secondary shorter cycle in HD 201092, although visual inspection indicates that this would be the case. HD 156206, on the other hand, does not have data to cover any cycle completely (from minimum to minimum). Whithout these stars, our sample consists of 18 stars, all 
with fairly clear cycles. All the stars in our analysis have also been found to be cyclic by \citet{olspert2018}.

Most of our stars are main-sequence stars, but we also include three giants. The MW database also includes \ion{Ca}{ii} H\&K measurements of the Sun. They were made by measuring the Moon because the lunar spectrum for the H\&K lines is effectively just reflected sunlight. Because the Mount Wilson data include only one full cycle for the Sun, we extended our data for the Sun by including Sacramento Peak (SP) \ion{Ca}{ii} K observations, which were scaled to the same level as the MW S-index as $S_{\rm{SP}} = 2.61 K_{\rm{SP}} - 0.0647$, as was done by \citet{olspert2018}. This combined data set includes three full solar cycles.

The series for the Sun was even further extended back to 1907, including data from solar cycles 15 to 24 by \citet{egeland2017}, who also added \ion{Ca}{ii} K plage index measurements from the Kodaikanal Observatory in India and calibrated them to the MW scale. However, we only used the data from MW and SP observatories, as was done in \citet{olspert2018}.

\begin{table*}
\caption{\label{minima} Our sample of Mount Wilson stars.} 
\centering 
\begin{tabular}{c c c c c c c c c c c c c}        
\hline\hline                 
  Star & $n_{\rm{cyc}}$ & $\log R'_{\rm{HK}}$ & $T_{\rm{eff}}$ [K] & $P_{\rm{rot}}$ [d] & $P_{\rm{cyc}}$ [yr] & $\langle \gamma \rangle$ & $\sigma$ & $\langle t_r \rangle /\langle t_d \rangle$ & $n_{\mathrm{bin}}$ & FAP & MS/G & Data \\    
\hline                        
  \object{HD 3651} & 1 & -5.040 & 5280 & 37.0 & 15.06 & 0.368 & ... & 0.594 & 10 & G & MS & 1995 \\      
  \object{HD 4628} & 2 & -4.874 & 5014 & 37.14 & 8.21$\pm$0.41 & 0.096 & 0.145 & 0.905 & 10 & E & MS & 1995 \\
  \object{HD 16160} & 1 & -4.902 & 4762 & 48.58 & 12.18 & 0.074 & ... & 1.023 & 10 & E & MS & 1995 \\
  \object{HD 26965} & 2 & -4.919 & 5196 & 38.65 & 10.34$\pm$0.07 & 0.076 & 0.131 & 0.864 & 10 & E & MS & 1995 \\
  \object{HD 32147} & 1 & -4.939 & 4703 & 33.7 & 10.40 & 0.127 & ... & 0.758 & 10 & E & MS & 1995 \\
  \object{HD 166620} & 1 & -4.975 & 5007 & 42.1 & 15.33 & 0.146 & ... & 0.723 & 10 & E & MS & 1995 \\
  \object{HD 219834A} & 2 & -5.098 & 5705 & 43.4 & 6.16$\pm$1.78 & 0.364 & 0.133 & 0.429 & 7 & G & G & 1995 \\
  \object{HD 219834B} & 3 & -4.919 & 5136 & 34.78 & 8.86$\pm$1.25 & 0.319 & 0.144 & 0.672 & 8 & E & G & 1995 \\
  \object{Sun} & 3 & -4.911 & 5780 & 26.09 & 10.90$\pm$1.23 & 0.394 & 0.161 & 0.625 & 10 & E & MS & 1995+SP \\
  \object{HD 10476} & 3 & -4.962 & 5489 & 35.6 & 10.23$\pm$0.16 & 0.111 & 0.273 & 0.697 & 10 & E & MS & 2001 \\
  \object{HD 81809} & 3 & -4.940 & 5889 & 41.66 & 7.94$\pm$0.59 & 0.222 & 0.074 & 0.740 & 9 & E & G & 2001 \\
  \object{HD 103095} & 3 & -4.939 & 5265 & 34.03 & 6.94$\pm$0.42 & 0.202 & 0.133 & 0.917 & 10 & E & MS & 2001 \\
  \object{HD 114710} & 3 & -4.738 & 6098 & 11.99 & 5.44$\pm$0.34 & 0.107 & 0.463 & 0.676 & 8 & G & MS & 2001 \\
  \object{HD 115404} & 2 & -4.502 & 4976 & 18.03 & 10.62$\pm$1.99 & 0.158 & 0.028 & 0.742 & 10 & G & MS & 2001 \\
  \object{HD 149661} & 6 & -4.625 & 5265 & 20.76 & 4.68$\pm$1.25 & -0.013 & 0.361 & 0.916 & 5 & G & MS & 2001 \\
  \object{HD 152391} & 3 & -4.469 & 5461 & 10.62 & 8.67$\pm$2.19 & -0.040 & 0.342 & 0.959 & 8 & E & MS & 2001 \\
  \object{HD 160346} & 4 & -4.818 & 4897 & 32.0 & 7.15$\pm$0.26 & 0.109 & 0.071 & 0.772 & 7 & E & MS & 2001 \\
  \object{HD 201091} & 4 & -4.588 & 4177 & 35.54 & 7.05$\pm$0.70 & -0.033 & 0.072 & 0.873 & 9 & E & MS & 2001 \\
  \hline                                   
\end{tabular}
\tablefoot{FAP = E/G (excellent/good) as defined by \citet{baliunas1995}. MS = main-sequence star, G = giant. $P_{\rm{cyc}}$ is given with its standard deviation; thus there are no \textquoteleft error bars for stars with only one detected cycle. $\gamma$ for the Sun was calculated from solar cycles 21-23, for which MW+SP data are available, but to calculate $\langle t_{\rm{r}}\rangle / \langle t_{\rm{d}}\rangle,$ cycles 1-20 were also included, where all $t_{\rm{min}}$ and $t_{\rm{max}}$ are from the dates listed in \citet{hathaway2015}. Source of $T_{\rm{eff}}$ for the Sun: \citet{Allens_Teff_sun}.}
\end{table*}

\subsection{Sunspot numbers}

To compare the stellar cycles to the solar cycle, we also analysed sunspot data in addition to the solar chromospheric measurements. We compared the MW+SP data to the classical Wolf sunspot number (WSN)\footnote{Source: WDC-SILSO, Royal Observatory of Belgium, Brussels; available at http://www.sidc.be/silso/datafiles} and to the group sunspot number (GSN), which is recalibrated for different observers with the active day fraction method by \citet{willamo}. Reaching as far back as 1610, the sunspot series is much longer than any time series of other active stars. We used the data for sunspot cycles 9-23, from 1843.5 to 2008.9, where multiple of these series are available (MW+SP, WSN, and GSN).

\section{Methods}

\subsection{Defining times of minima and maxima}

We defined the times for minima and maxima of the stellar activity cycles individually for each cycle. To define the exact time, we fitted a parabola to the data around the minimum or maximum, and the interval of data included varied depending on the specifics of the cycle. When the cycle was very asymmetric around the minimum or maximum, only a short interval could be used when a symmetric function was fitted, whereas with a poorly covered cycle, a longer interval had to be used to obtain enough data for a reliable fit. The times of minima and maxima defined by this method along with the intervals we used are listed in the appendix (Table \ref{append1}). One cycle is then defined as the time between two consecutive minima.

For the dates of minima and maxima for the Sun, we used the minimum and maximum value of the 13-month mean value of the sunspot number. This is a commonly used definition of solar minima \citep[see e.g.][]{hathaway2015}. This is the minimum of the sunspot number cycle, and the chromospheric emission need not be at its minimum at the same time -- there are indeed differences of even several years in the timing of the solar minima between different activity indicators, such as the sunspot number, sunspot area, and 10.7cm radio flux \citep{hathaway2015}. When the same minima times are used for different solar activity indicators, however, the analysis for the MW cycles of the Sun is comparable to that for the sunspot cycles in Section \ref{sunspotcycle}. For other stars we have only MW data, therefore they are not directly comparable in this sense to the solar cycle.

\subsection{Skewness} \label{skewness}

Skewness is a statistical measure of the asymmetry of a probability distribution, which has been used to measure asymmetries of solar cycles \citep{ramaswamy1977,lantos2006,du2011}. The skewness $\gamma$, or third moment, of a set of data points $x_{i}$ is defined as

\begin{equation}
\gamma = \frac{\sum\limits_{i=1}^{N} (x_i - \overline{x})^3}{(N-1)\sigma_x^3},
\end{equation}

\noindent where $N$ is the number of data points, $\overline{x}$ is the sample mean, and $\sigma_x$ is the standard deviation of the sample. A positive skewness indicates a distribution leaning to the left, or in the case of a stellar cycle, a cycle with faster rise time and slower declining time. A negative skewness indicates a leaning to the right, or longer rise time and shorter declining time. A symmetric distribution has $\gamma=0$, although zero skewness does not always mean that the distribution is symmetric. For instance, a distribution with a long and thin tail on the one side and short but thick on the other could also have $\gamma=0$.

In order to calculate the skewness of an activity cycle, the cycle has to be transformed into a one-dimensional distribution. We did this by dividing the cycle into ten bins of equal length, where the centre of the bin is at $t_{\rm{bin}}$. In the cases when the gaps in the data were too long and some bins would have no data points at all, we reduced the number of bins into the largest number that still included data points in each bin. 
Then we calculated the mean value of the data points in each bin, and built the final distribution, emulating the cycle, by multiplying this mean value by 10 000 in order to derive an integer value $n$ from data with four decimals, and added $n$ occurrences of $t_{\rm{bin}}$ to the distribution.

\begin{figure}
   \centering
   \includegraphics[width=\hsize]{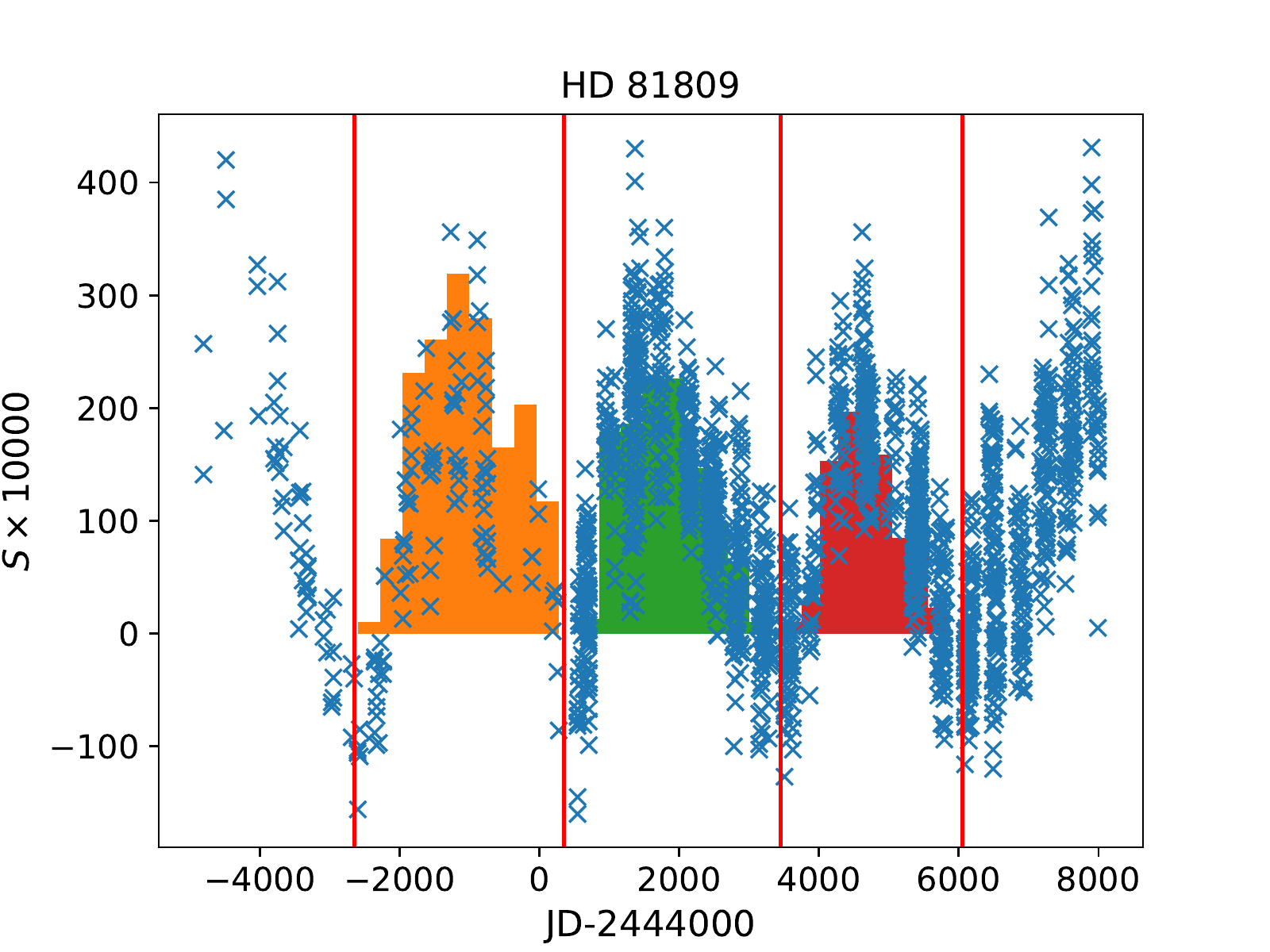}
   \caption{Cycles of HD 81809. The crosses are the original calibrated MW data, and the histograms are the distributions built from these. Vertical lines show the times of minima, dividing the data set into three complete cycles. The zero-levels of the histograms are defined individually for each cycle, but here they are plotted on the same level. The correct individual shifts for each cycle are therefore missing in the visualisation for simplicity. The value on the y-axis, $S \times 10000$ (which has been shifted in the y-direction), equals the number of data points in a bin, $n$.}
    \label{81809}
\end{figure}

In order to compare the skewness of stellar cycles to the solar cycle, their zero-levels must be comparable. The sunspot cycle approaches zero at solar minimum, but the S-index of active stars does not. To correct for this, we shifted all the bins of a cycle with a constant value, so that the bin with the lowest value reached $S_{\rm{min}}=0.001$ (corresponds to the $t_{\rm{bin}}$ appear $n=10$ times in the distribution). This was performed similarly for each cycle. An example of this type of distributions emulating the cycles of HD 81809 is shown in Fig. \ref{81809}. We repeated the same analysis with the same shift of the zero-level also for the sunspot cycle.
For each star, we calculated the skewness for each cycle, and used the average of these cycles as a measure of the average cycle asymmetry for this star.

\section{Results}

\subsection{Rise and decline times of cycles}

A simple way to estimate the asymmetry of a cycle is to compare the duration of the rising and the declining phases of the cycle. In the Sun the rising phase is typically shorter.

\begin{figure}
   \centering
   \includegraphics[width=\hsize]{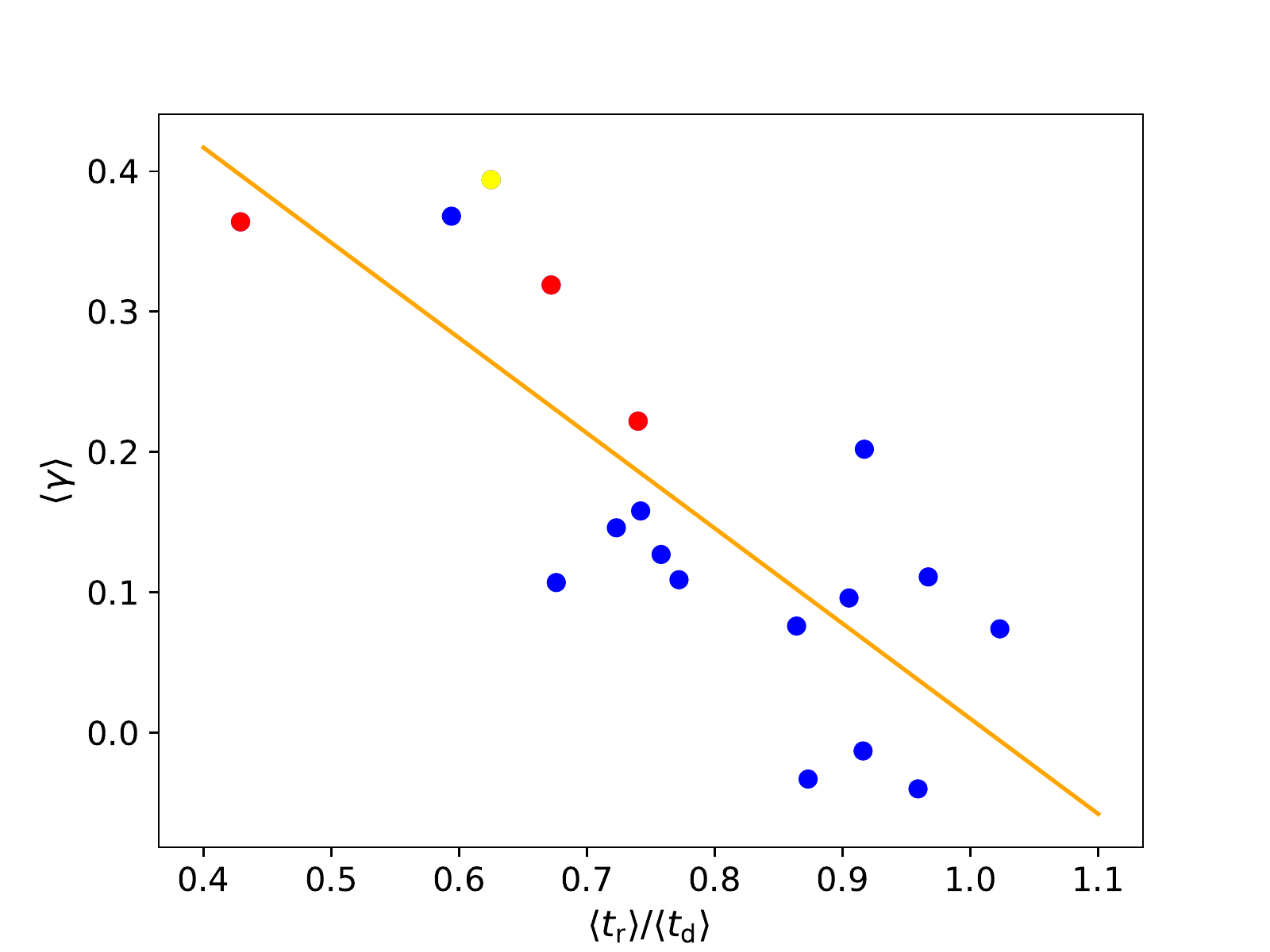}
   \caption{Ratio of average rise time and average decline time vs. average skewness. Blue dots represent main-sequence stars, and red dots giants. The Sun is shown in yellow. The continuous line shows the best linear fit.}
    \label{tr_td}
\end{figure}

For each star we calculated the ratio of the average duration for the rising phase $\langle t_{\rm{r}}\rangle$ and average duration of the declining phase $\langle t_{\rm{d}}\rangle$ of a cycle. Figure \ref{tr_td} shows the relation of this ratio to the average skewness of the stellar cycles. As both are a measure of asymmetry, the almost linear relation is expected. The values of $\langle t_{\rm{r}}\rangle/\langle t_{\rm{d}}\rangle$ are also listed in Table \ref{minima}.

For the calculation of the $\langle t_{\rm{r}}\rangle/\langle t_{\rm{d}}\rangle$ parameter for the Sun we used sunspot data for solar cycles 1-23 for better statistics. With any other star, the maximum number of cycles is six.

As the main measure of the cycle asymmetry we used the skewness of the cycle (see Sect. \ref{avg_skew}), but the correlation of the skewness and $\langle t_{\rm{r}}\rangle / \langle t_{\rm{d}}\rangle$ confirms that both are usable parameters to measure cycle asymmetries. We calculated a Pearson correlation coefficient $r = -0.78$ between these two parameters, which indicates a fairly strong negative correlation. Assuming linearity, we derived the relation between them as

\begin{equation}
  \gamma=-0.68 \Big[\langle t_{\rm{r}}\rangle / \langle t_{\rm{d}}\rangle \Big] + 0.69. 
\end{equation}

\noindent The times of minima and maxima are listed for each star in the appendix (Table \ref{append1}). When available, we used times of maxima of incomplete cycles to obtain better statistics. For instance, for HD 32147  only one complete cycle is available, but three times of maxima.

\subsection{Average skewness of MW cycles} \label{avg_skew}

\begin{figure}
   \centering
   \includegraphics[width=\hsize]{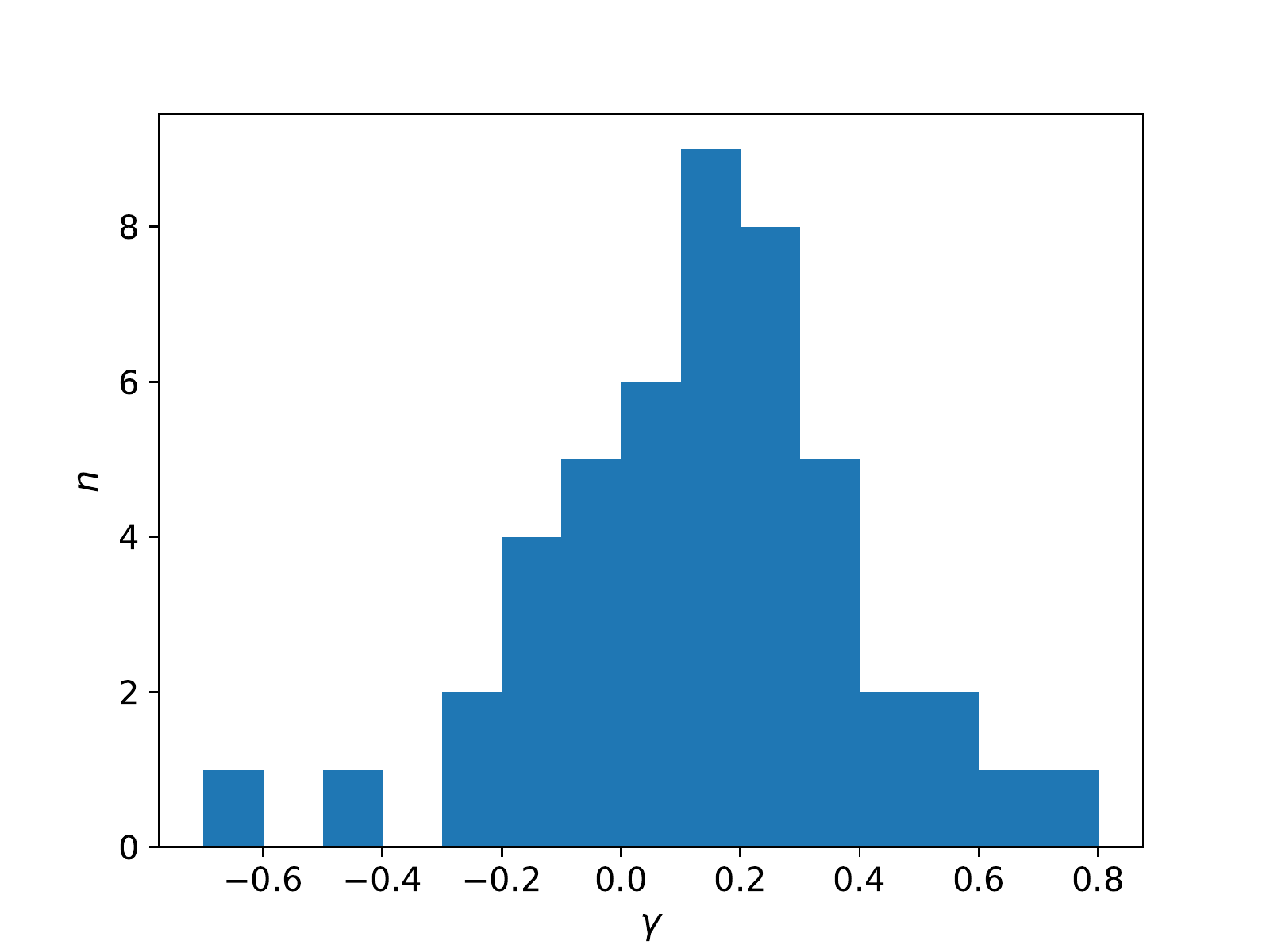}
   \caption{Distribution of skews of all cycles of all stars.}
    \label{S_distr}
\end{figure}

The average skewness $\langle \gamma \rangle$ for the cycles of each star and its standard deviation $\sigma$ for those stars with multiple cycles is shown in Table \ref{minima}.
Figure \ref{S_distr} shows the distribution of the skews of all cycles of all stars. Most cycles (34 of a total number of 47 = 72\%) have a positive skew. The peak values are between 0.1 and 0.2. The Sun has a considerably high asymmetry, with a mean skew from MW+SP data of 0.394. 
Taking all 47 cycles into account, we obtain an average skewness of 0.13, with a standard deviation 0.26.

\begin{figure}
   \centering
   \includegraphics[width=\hsize]{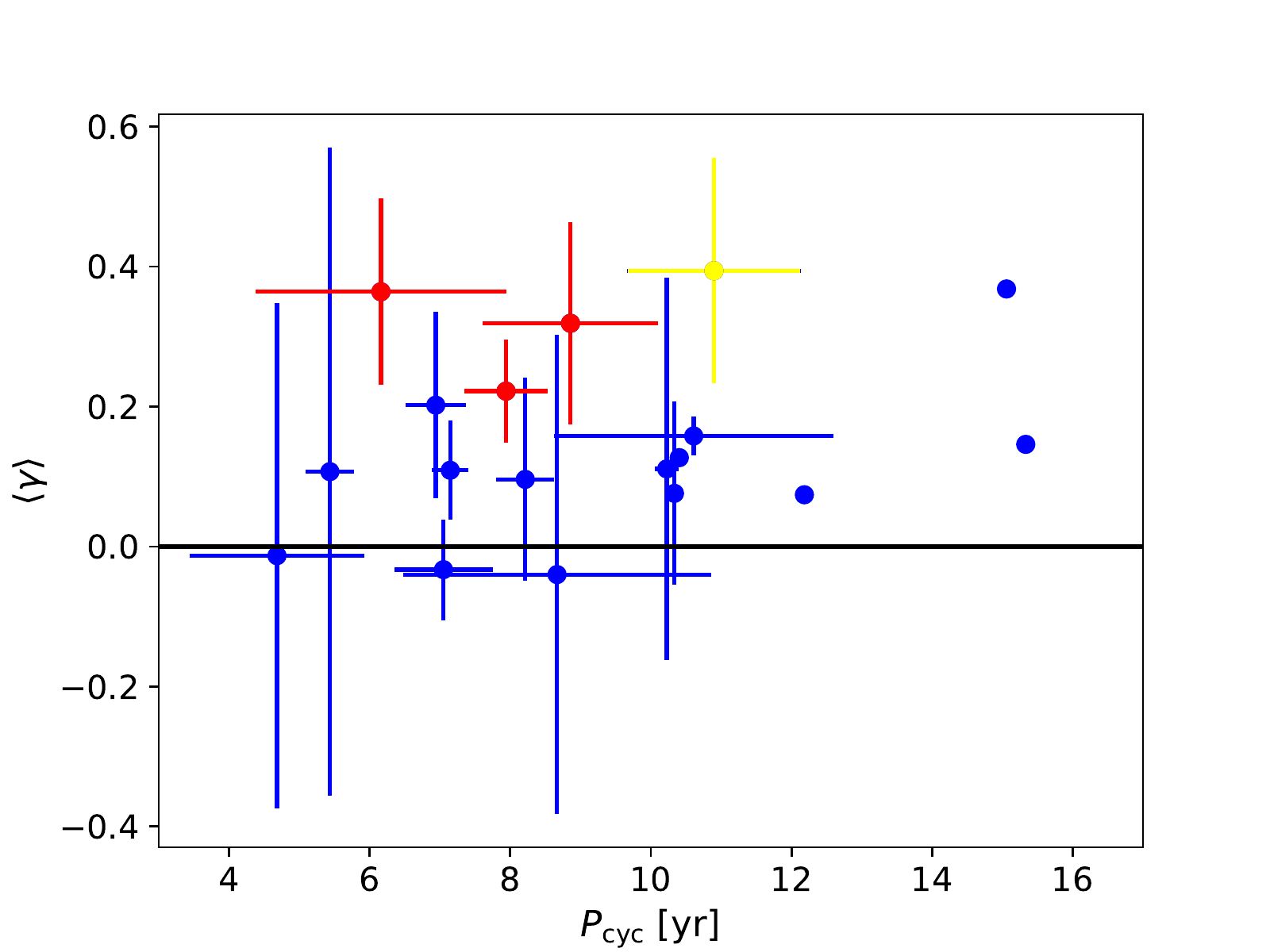}
   \caption{Average skewness plotted against $P_{\rm{cyc}}$. Blue dots represent main-sequence stars, and red dots giants. The Sun is shown in yellow. The error bars represent the cycle-to-cycle variations for stars with multiple cycles. The vertical line represents $\gamma = 0$.}
    \label{S_Pcyc}
\end{figure}

\begin{figure}
   \centering
   \includegraphics[width=\hsize]{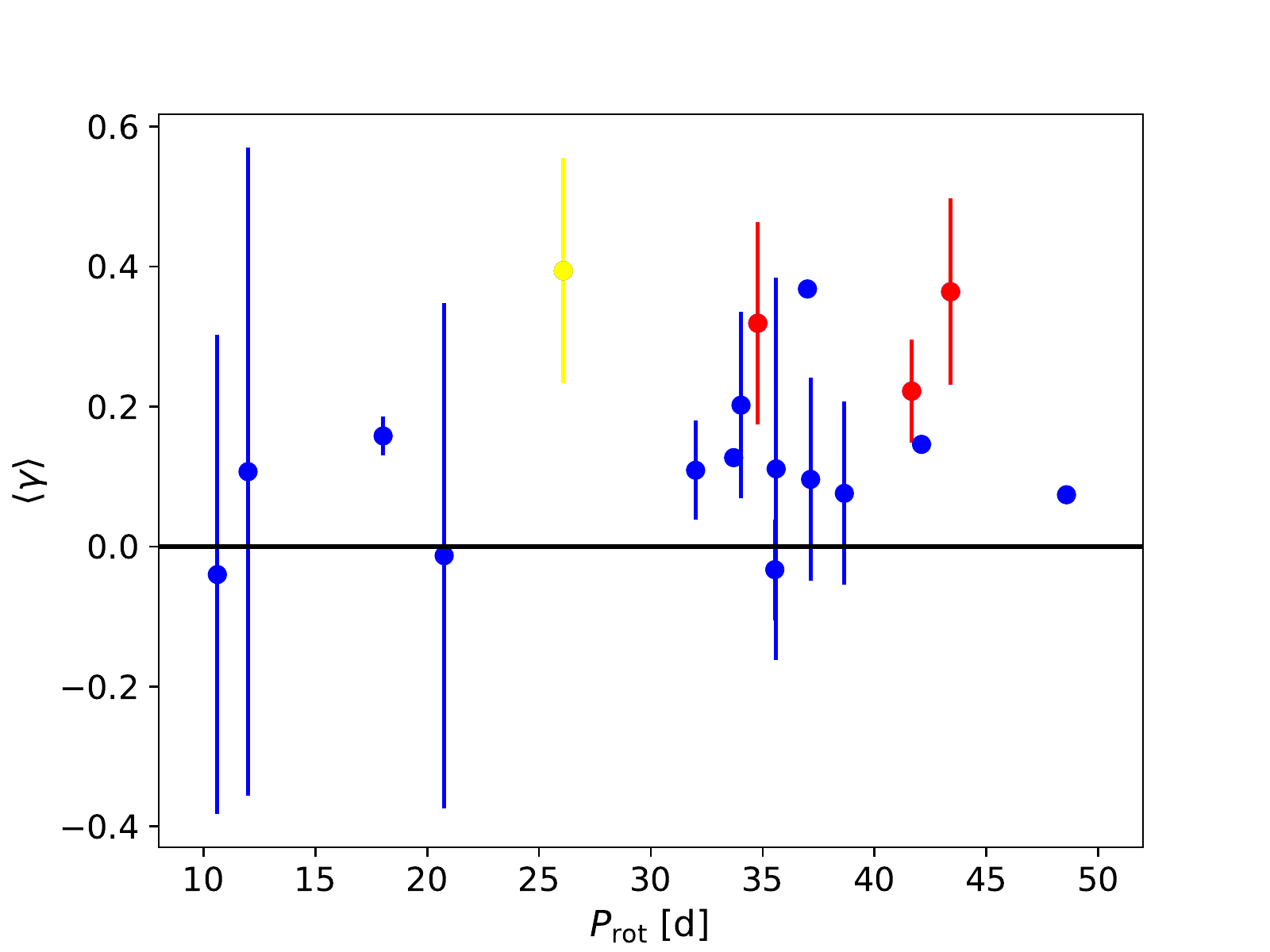}
   \caption{Same as Fig. \ref{S_Pcyc}, but for $P_{\rm{rot}}$.}
    \label{S_Prot}
\end{figure}

\begin{figure}
   \centering
   \includegraphics[width=\hsize]{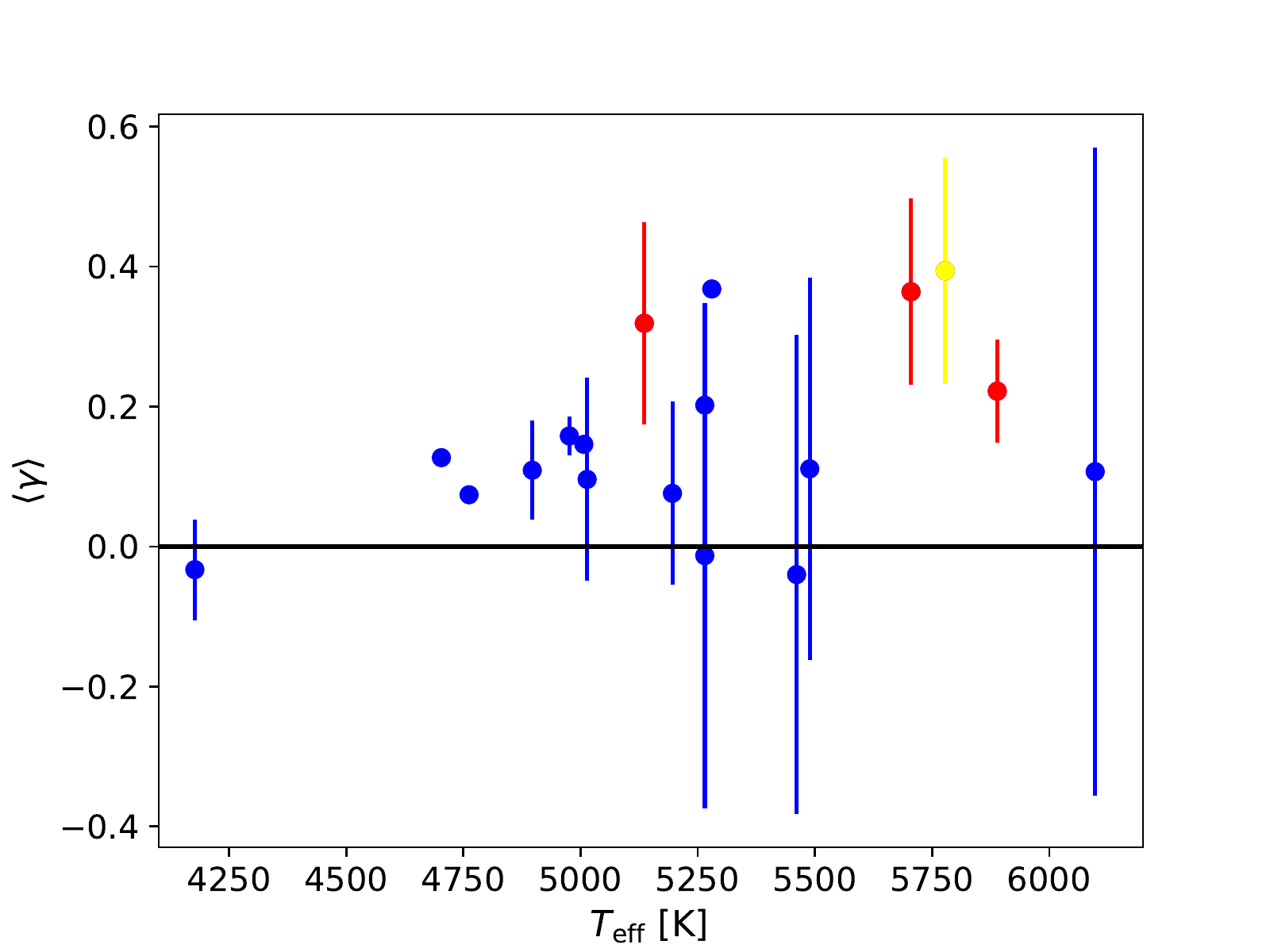}
   \caption{Same as Fig. \ref{S_Pcyc}, but for $T_{\rm{eff}}$.}
    \label{S_Teff}
\end{figure}

\begin{figure}
   \centering
   \includegraphics[width=\hsize]{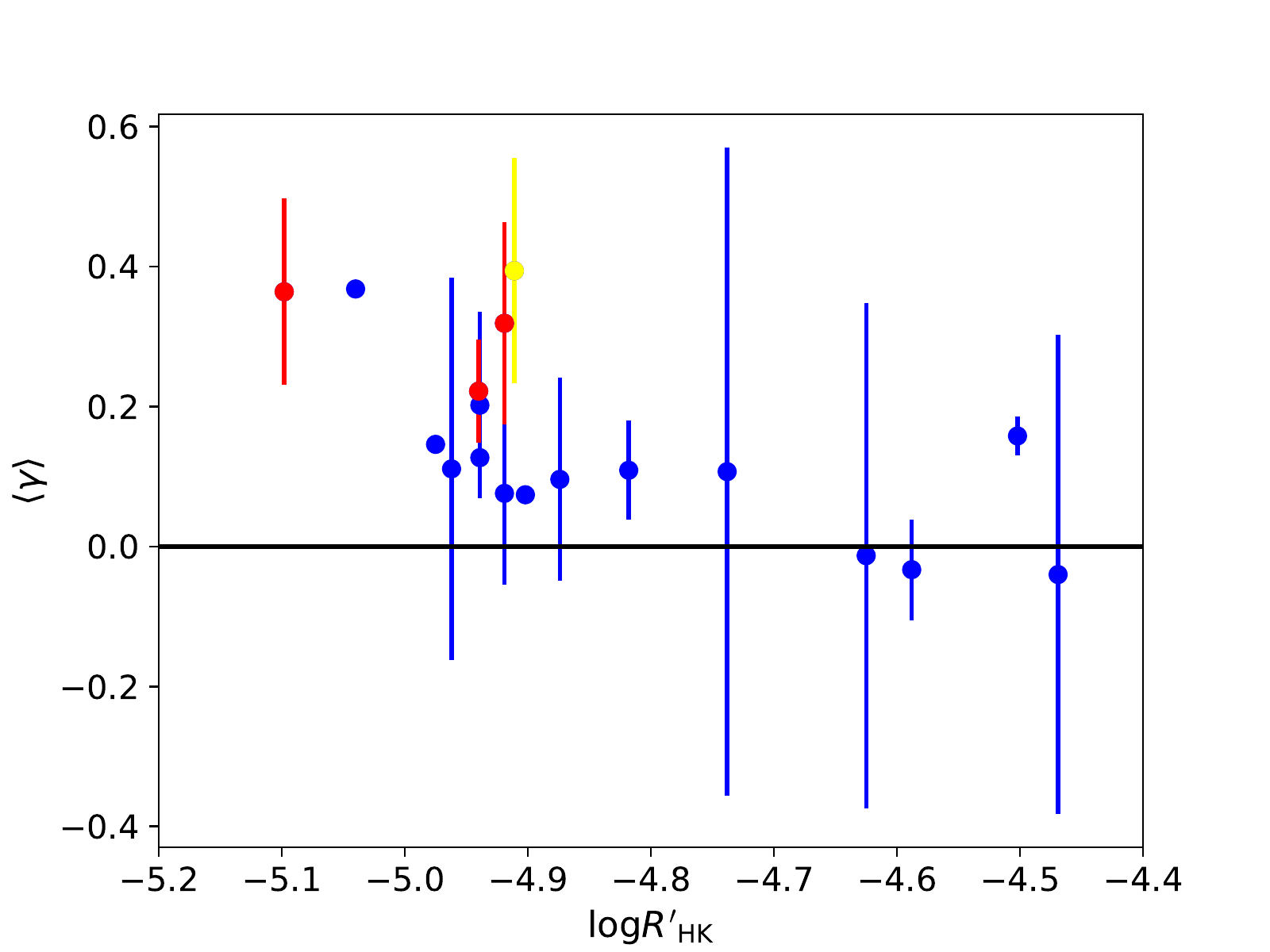}
   \caption{Same as Fig. \ref{S_Pcyc}, but for $\log R'_{\rm{HK}}$.}
    \label{S_RHK}
\end{figure}

We compared the average skewness for each star to other stellar parameters; cycle period $P_{\rm{cyc}}$, rotation period $P_{\rm{rot}}$, effective surface temperature $T_{\rm{eff}}$ , and activity index $\log{R'_{\rm{HK}}}$, in Figs. \ref{S_Pcyc}-\ref{S_RHK}. The figures show the mean value and standard deviation of the variation of $\gamma$ (and $P_{\rm{cyc}}$) for stars with multiple cycles ($\sigma$ and the standard deviation of $P_{\rm{cyc}}$ in Table \ref{minima}). Values for $P_{\rm{rot}}$ and $\log R'_{\rm{HK}}$ are from \citet{olspert2018}. The $T_{\rm{eff}}$ values are from Gaia DR2 \citep{Gaia2016,GaiaDR2,Gaia2018_Teff}, except for the Sun.

The three giants and the Sun have a considerably high skewness; the Sun has the highest skewness of the stars in our sample. 
This is mainly due to the third cycle (solar cycle 23), which is the second most positively skewed cycle of any star in our sample; solar cycles 21 and 22 are much more symmetric. Sunspot data also give a much lower skewness for cycle 23 than MW+SP data (see Sect. \ref{sunspotcycle}). The value $\langle t_{\rm{r}}\rangle/\langle t_{\rm{d}}\rangle = 0.625$ for the Sun, which was calculated from sunspot cycles 1-23, is also a very asymmetric value, but  two stars have an even larger asymmetry in the rise and decline times. The skewness of the solar cycles from MW+SP data might thus be slightly biased as a result of an over-representation of very asymmetric cycles. We recall, however, that the rise and decline times were calculated from sunspot data, which might behave differently than chromospheric data.

We calculated the Pearsons correlation coefficients $r$ between $\langle \gamma \rangle$ and the other parameters. These are shown in Table \ref{corr} along with their $p$-values.  $\langle \gamma \rangle$ and $P_{\rm{cyc}}$ or $P_{\rm{rot}}$ show at best a very weak positive correlation. There might be a slightly stronger positive correlation between $\langle \gamma \rangle$ and $T_{\rm{eff}}$, but the most relevant is the negative correlation between $\langle \gamma \rangle$ and $\log R'_{\rm{HK}}$ ($r = -0.67$). The less active stars might thus have more asymmetric cycles in general. This is plausible because young, active stars are known to have more irregular cycles than older, less active stars \citep{baliunas1995}. Irregular cycles might be skewed in either direction and then be averaged close to symmetric cycles with zero skewness, if enough cycles are included. This correlation is unclear, however. More data are required to be analysed before this can be claimed with some certainty. This differs from the simulated results of \cite{pipin2016}, who found stronger cycles to be more asymmetric in the regime of weak cycles in their mean-field simulations.

There is much variation in the skewness of the cycles for individual stars, which is seen as high values of $\sigma$. Table \ref{minima} shows that $\sigma$ is generally high compared to $\gamma$. This is probably not only due to the limited amount of cycles because the sole star with 6 cycles (HD 149661) has the second highest value of $\sigma$. The large cycle-to-cycle variations are expected because this is the case in the Sun as well (see Sect. \ref{sunspotcycle}).

\begin{figure}
  \centering
  \includegraphics[width=\hsize]{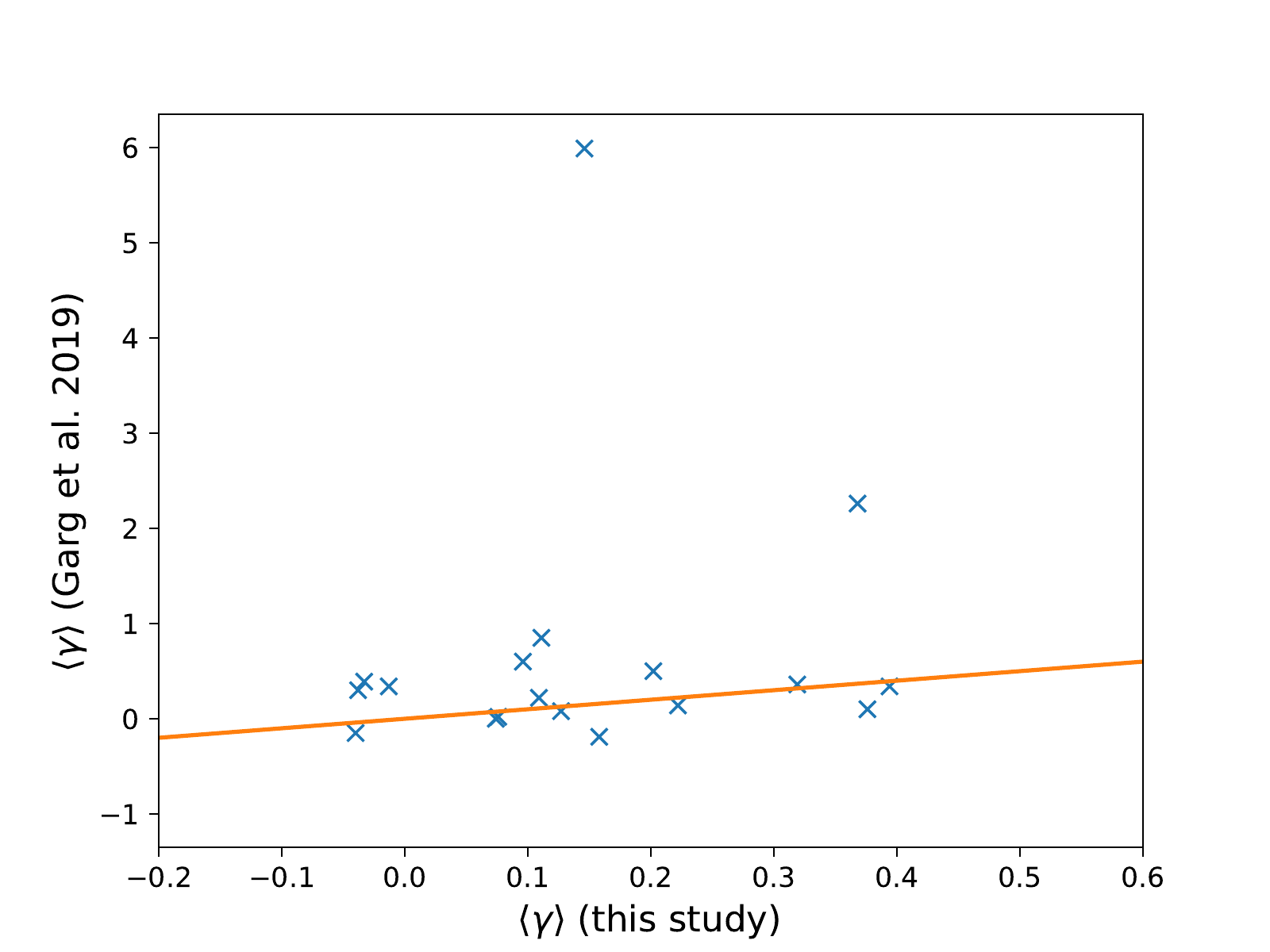}
  \caption{Our average skewness for each star compared to that of \citet{garg2019}. The orange line is y=x, where these values would be equal.}
  \label{garg}
\end{figure}

We also compared our values for the average skewness to those of \citet{garg2019}, who studied the same data. This is shown in Fig. \ref{garg}. There are some large differences in the values. This might be due to the definition of the zero-level or the binning, which are not described in detail in \citet{garg2019} because they focused more on the Waldmeier effect than on the cycle skews.

One source of uncertainty is the number of bins, which might affect the skewness of the cycle. In most cases, the data are abundant and regular enough to allow us to divide them into ten bins, but in some stars the number of bins is reduced, in the worst case, to five bins. This is inevitable because these data contain large gaps. The number of bins used for each star is listed in Table \ref{minima}.

We tested the effect of the binning with the Sun, for which $n_{\rm{bin}} = 10$, and average skewness $\langle \gamma \rangle = 0.394$. When the number of bins is reduced to $n_{\rm{bin}} = 8,$ we obtain $\langle \gamma \rangle= 0.402$, and with $n_{\rm{bin}} = 5$, $\langle \gamma \rangle= 0.329$. For stars with poorer data quality, this effect can be expected to be even stronger. It appears to be evident, however, that the stars with $n_{\rm{bin}} = 10$ are most comparable to each other.

Because of the gaps in the data, long cycles are probably more reliable than short cycles because the seasonal gaps affect a proportionally smaller part of the cycle, and the shape of the cycle can be identified with more certainty.

\begin{table}
\caption{Correlation coefficients between $\gamma$ and other parameters.}             
\label{corr}      
\centering                          
\begin{tabular}{c c c}        
\hline\hline                 
Parameter & Correlation coefficient & $p$-value \\   
\hline                        
$\langle t_{\mathrm{r}}\rangle / \langle t_{\mathrm{d}}\rangle$ & -0.78 & $1.3 \times 10^{-4}$ \\ 
$P_{\mathrm{cyc}}$ & 0.28 & 0.26 \\ 
$P_{\mathrm{rot}}$ & 0.29 & 0.24 \\ 
$T_{\mathrm{eff}}$ & 0.42 & 0.08 \\ 
$\log R'_{\mathrm{HK}}$ & -0.67 & $2.6 \times 10^{-3}$ \\ 
\hline
\end{tabular}
\end{table}

\subsection{Average cycle shape}

We also tried to combine all 47 individual cycles of all stars into an average cycle. The cycle amplitudes were normalised with the same binning as was used in the calculation of $\gamma$, so that the lowest bin has the value 0 and the highest bin 1. We added all the data points from individual cycles, scaled between 0 and 1, to the combined cycle without any averaging. The scaling of individual cycles was, however, made with the mean values of the bins to avoid that extreme data points set the scale for the cycles. The cycle duration was normalised to a phase between 0 and 1.

For the resulting average cycle, $\gamma$ was calculated similarly as for individual cycles, except that now we divided the data to 20 bins instead of 10 because the data are much more abundant. We obtain the value for the skewness as $\gamma=0.078$, which is slightly lower than the average skewness for individual cycles.

\begin{figure}
   \centering
   \includegraphics[width=\hsize]{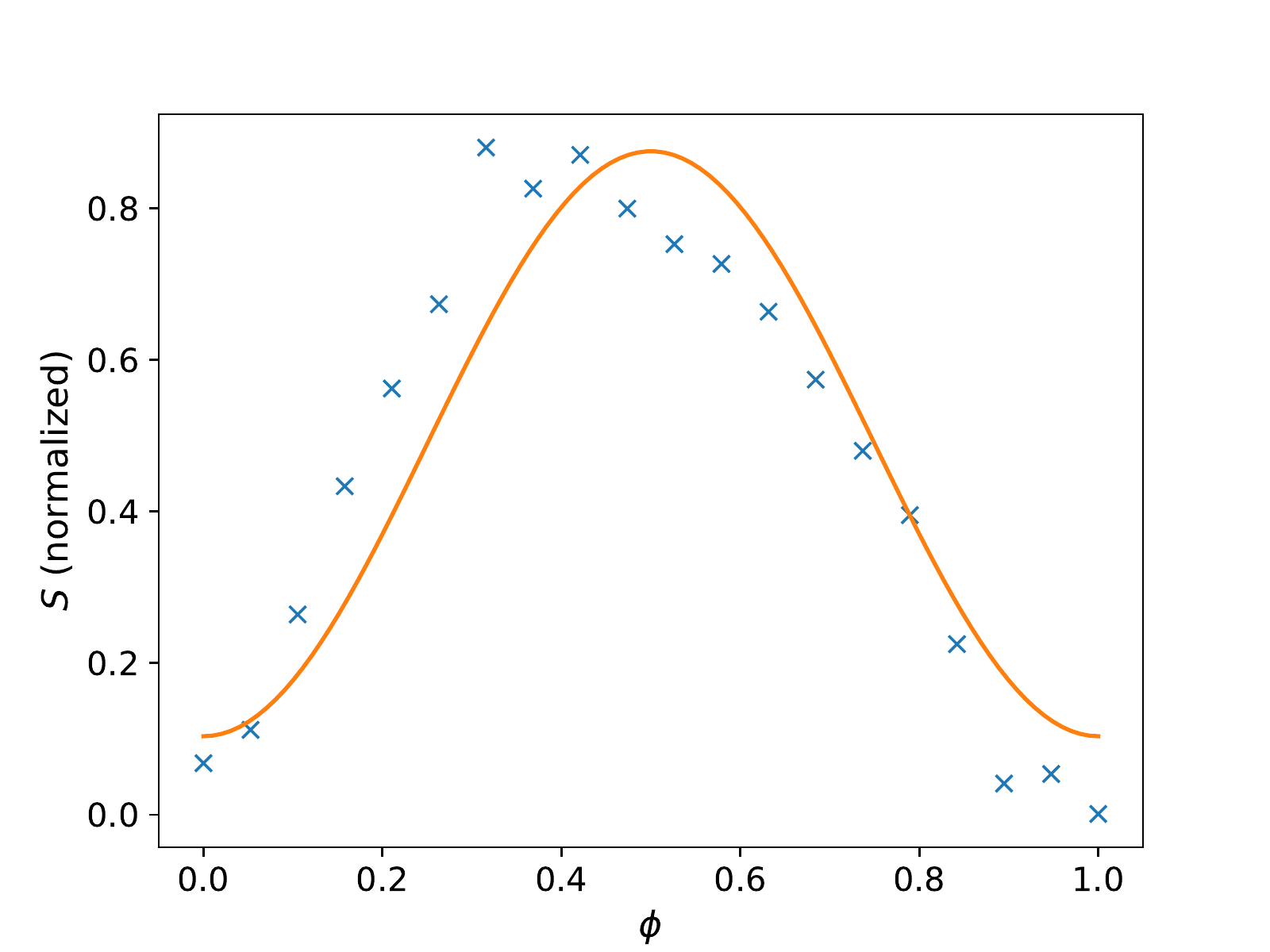}
   \caption{All cycles phased and normalised together. The blue crosses show the resulting cycle, and the orange curve is a cosine fitted to the cycle.}
    \label{all_cycles}
\end{figure}

We fitted a sinusoid of the form

\begin{equation}
S(\phi)=a\cos(2\pi \phi)+c
\end{equation}

\noindent to the averaged cycle with 20 data points (same as in the binning when we calculated $\gamma$). The cosine function has its minimum or maximum at 0, which is defined as the cycle minimum, so that it is forced to the same phase as the cycle. The fitted cosine function, with the fitted parameters $a=-0.386$ and $c=0.489$, is shown in Fig. \ref{all_cycles}. The fit is plausible, even though individual cycles can be very irregular. Quantitatively, we obtain the chi-squared statistics between the data points and the fitted sinusoid as $\chi^2 = 11.43$, and the $p$-value $p = 0.91$. 
The cosine curve, however, is not able to take the asymmetry into account because the actual cycle rises to its maxium faster than the cosine. The best-fit cosine also has its maximum at a similar level as the actual average cycle, but its minimum is not as deep. This feature is different than the feature described by \citet{reinhold17} for Kepler stars, where the maximum was sharper, and the minimum flatter than the sine curve. They used the amount of photometric variability as a proxy of magnetic activity, however. The variability should be highest around activity maximum, but the details of the cycles might still be different than those found from the S-index, and furthermore, the span of the Kepler data allow the detection of cycle periods only up to around six years.

In Fig. \ref{all_cycles} there might also be an indication of a double peak, as is commonly seen in the Sun, with the Gnevyshev gap in between. The feature is rather weak, however, therefore based on our data, we do not claim the existence of double peaks and the Gnevyshev gap in other stars.

\subsection{Comparison to sunspot cycles} \label{sunspotcycle}

The stellar data can be compared to sunspot data. We used the same method to calculate the skewness for monthly values of both the classical WSN and the group sunspot number (GSN) series. The GSN ignores individual spots and only counts the number of spot groups, which reduces observational errors and makes the old observations more reliable. We used the minimum values of the 13-month average number of sunspots as the times of solar minima: 1843.5, 1855.9, 1867.2, 1878.9, 1890.2, 1902.0, 1913.5, 1923.6, 1933.7, 1944.1, 1954.3, 1964.8, 1976.2, 1986.7, 1996.3, and 2008.9 (see e.g. Hathaway 2015).

Our values for the skewness of solar cycles 9 to 23 are shown in Table \ref{GSN}. We also compare our values to the skewness for the WSN published by other authors. Our values agree well with those of \citet{lantos2006}, but not so well with those of \citet{du2011}. The skewness of the GSN is very similar to the skewness of the classical WSN.

The are some notable differences in the MW data and sunspot data. Especially for cycle 23 do the MW data give a very high skewness of $\gamma=0.614$, whereas sunspot data give $\gamma=0.282$. If the MW cycles for the Sun are not comparable to the sunspot cycle, then cycles for other stars cannot be expected to be directly comparable to the sunspot cycle either.

\begin{table}
\caption{Skewness of the solar cycles.}             
\label{GSN}      
\centering                          
\begin{tabular}{c c c c c c}        
\hline\hline                 
\# & MW+SP & WSN & GSN & \citet{lantos2006} & \citet{du2011} \\    
\hline                        
  9 & \ldots & 0.114 & -0.004 & 0.235 & 0.507 \\      
  10 & \ldots & 0.400 & 0.389 & 0.346 & 0.135 \\
  11 & \ldots & 0.565 & 0.490 & 0.646 & 0.522 \\
  12 & \ldots & 0.468 & 0.360 & 0.414 & 0.087 \\
  13 & \ldots & 0.559 & 0.525 & 0.640 & 0.345 \\
  14 & \ldots & 0.086 & 0.079 & 0.204 & -0.074 \\
  15 & \ldots & 0.339 & 0.342 & 0.314 & 0.327 \\
  16 & \ldots & 0.180 & 0.191 & 0.262 & 0.020 \\
  17 & \ldots & 0.417 & 0.425 & 0.299 & 0.122 \\
  18 & \ldots & 0.294 & 0.264 & 0.273 & 0.162 \\
  19 & \ldots & 0.629 & 0.607 & 0.581 & 0.299 \\
  20 & \ldots & 0.375 & 0.232 & 0.330 & 0.043 \\
  21 & 0.338 & 0.134 & 0.158 & 0.299 & 0.116 \\
  22 & 0.231 & 0.357 & \ldots & 0.419 & 0.164 \\
  23 & 0.614 & 0.282 & \ldots & \ldots & 0.300 \\
\hline
\end{tabular}
\end{table}

\section{Comparison to simulations}

To compare our observational results to numerical simulations, we used the direct numerical magnetohydrodynamic (MHD) simulations of convective dynamos in solar-like stars, described in \citet{viviani2018}, \citet{warnecke2018}, and \citet{warnecke2019arxiv}. Some of the simulations, presented in \cite{viviani2018}, are global MHD simulations, ranging between 0.7-1.0 $R$ in the radial direction, and only omitting the polar regions, modelling the star between latitudes -75$^\circ$ to +75$^\circ$, and the full longitudinal range. A few of the runs in \citet{viviani2018} and all the runs in \citet{warnecke2018} and \citet{warnecke2019arxiv} are wedges in the azimuthal direction, covering only the longitudes from 0 to $\pi/2$. These are labelled with the superscript \textquoteleft W' in Table \ref{mariangela}. A few of the global simulations are run in higher resolution. These are marked with the superscript \textquoteleft a'. The higher resolution runs are slightly more realistic because they are more turbulent than their lower resolution counterparts. In addition to comparing  them with the observed MW cycles, we also investigated whether the differing geometry of the simulation setup affects the results.

In all of these runs, turbulent convection under the influence of rotation generates differential rotation and large-scale dynamo action. As a result, dynamically significant dynamo modes at the system scale are generated and maintained by the flow.

The radial magnetic field at 0.98 $R$ is decomposed into spherical harmonics, where $m=0$ mode contains the axisymmetric part of the radial magnetic field, $m=1$ is the first non-axisymmetric mode, $m=2$ is the second mode, and so on. We studied the evolution of the dominating dynamo mode in each simulation (found in Table 4 in \citet{viviani2018}), which is $m=0$ or $m=1$ in all runs. In all the wedge runs $m=0$ is the dominating mode, containing most of the magnetic energy on large scales. We note here that a substantial amount of magnetic energy in all runs comes from the small-scale non-axisymmetric field, but for the comparison with observed cycles, only the large-scale magnetic field is relevant.

We chose the runs where cycles for the dominating mode could be defined for a closer study; this includes 20 runs in total. 
We chose only runs where more than one cycle could be identified in order to obtain some estimate for the cycle-to-cycle variability. 
We point out that the simulations do not always produce strictly cyclic dynamo solutions, which is likely due to the competition of different dynamo modes in the simulated system. Therefore defining the cycle minima was more challenging from the models than from the MW data. Thus, the results may not be as reliable for the simulated data.

We built the distributions emulating the cycles similarly as for the MW data by multiplying the value of each data point so that we obtained an integer number, and added this many occurrences of this time point to the distribution. We then fitted a parabola around the minimum to define its exact location. Then we calculated the skewness of each cycle similarly as with the MW data.

\begin{figure*}
   \centering
   \includegraphics[width=6cm]{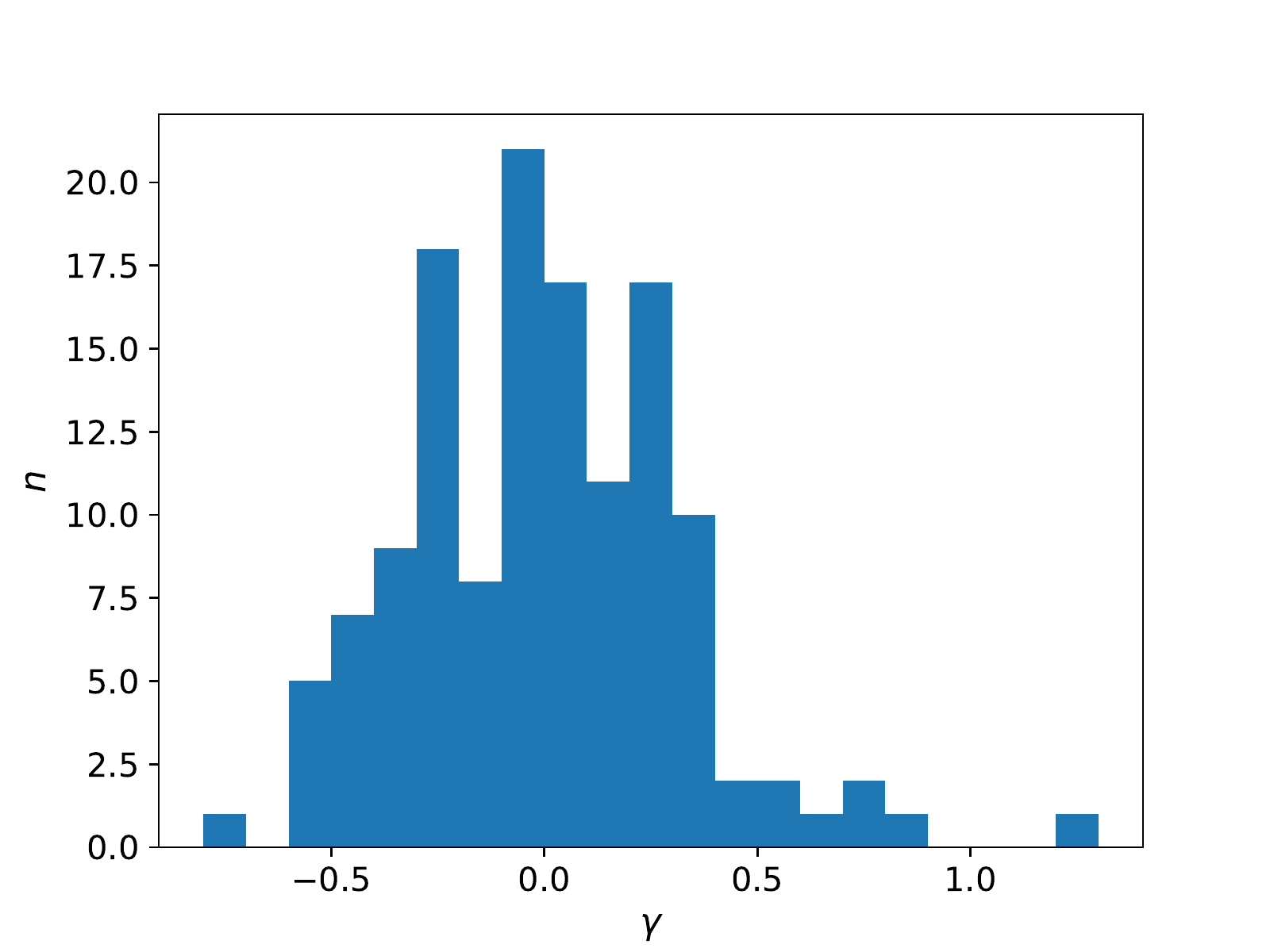}
   \includegraphics[width=6cm]{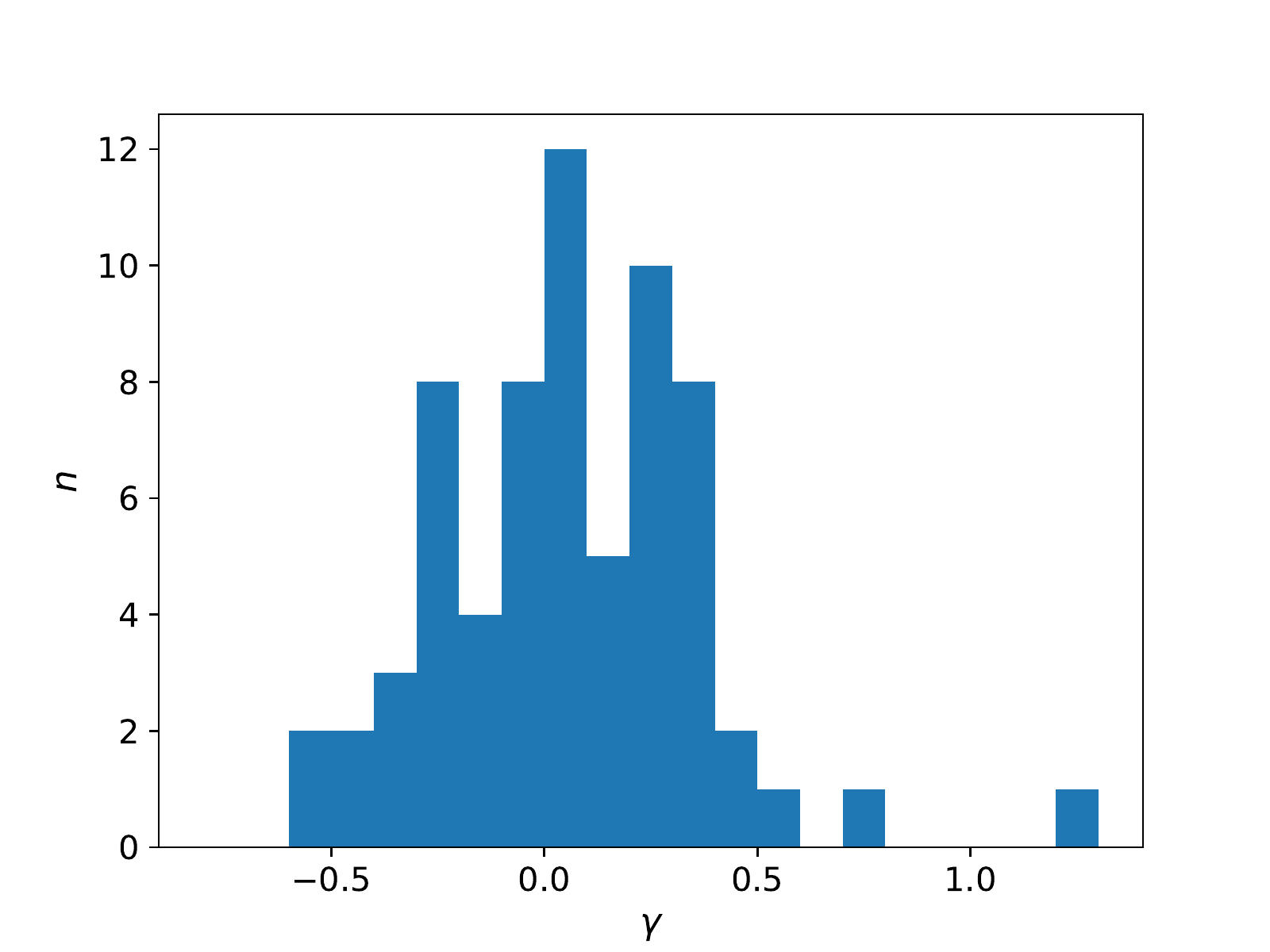}
   \includegraphics[width=6cm]{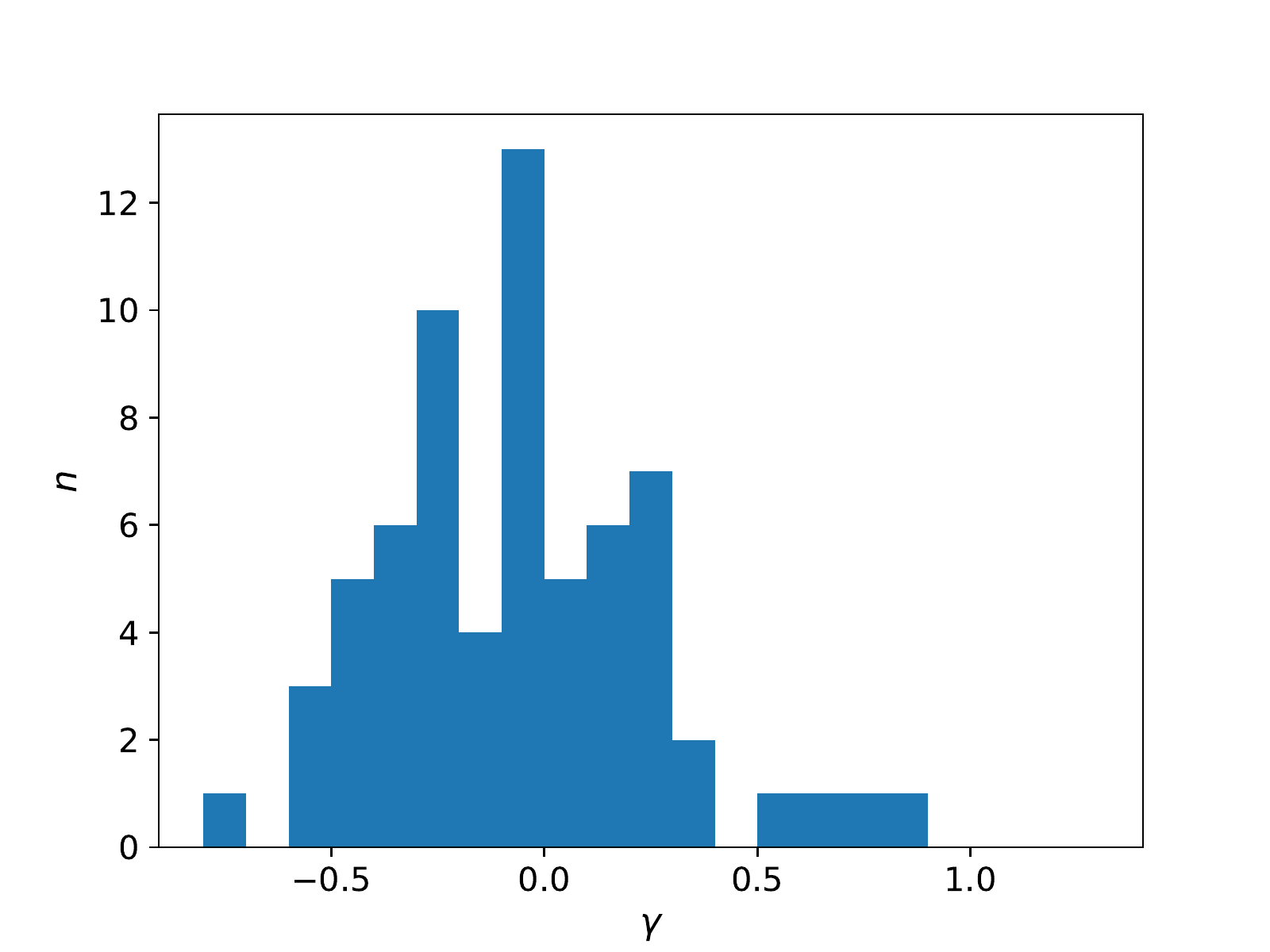}
   \caption{Distribution of skews of individual cycles for all the runs (left), only the global runs (centre), and only the wedge runs (right).}
   \label{Mariangela_hist}
\end{figure*}

Figure \ref{Mariangela_hist} shows a histogram of the distribution of the skews of the simulated cycles for all cycles together and including only the global runs or only the wedge runs. There is a visible difference between the global and wedge runs: while the histogram including all cycles is centred around zero, with a mean skewness 0.00 and standard deviation 0.32, the histogram including only the global runs has a mean skewness of 0.06 and standard deviation 0.31, and the one including only wedge runs has a mean of -0.06 and standard deviation 0.31. It would thus seem that global simulations produce more positively skewed cycles than wedge runs, although in both cases the cycle-to-cycle variation is large, as it is in real stars as well.

In both global and wedge runs, the deviation (0.31 in both cases) is larger than the difference between them ($0.06 - (-0.06) = 0.12$). To investigate if the difference is significant, we additionally calculated the standard error $\sigma_{\langle \gamma \rangle}$ of the mean of the distribution:

\begin{equation}
  \sigma_{\langle \gamma \rangle} = \frac{\sigma^2}{n},
\end{equation}

\noindent where $n$ is the sample size. We obtain $\sigma_{\langle \gamma \rangle, \rm{global}} = 0.04$ and $\sigma_{\langle \gamma \rangle, \rm{wedge}} = 0.03$. These are smaller than the difference, which indicates that it is significant and not noise caused by a small sample size. However, we note that for the global runs, a significantly large fraction of the cycles (14 of 48) are from the run K1, which has higher average skewness than most of the runs, and might induce a bias to the result. Nevertheless, the difference between the global and wedge runs, although small, is probably real.

The wedge assumption forces the large-scale dynamo to be axisymmetric, whereas in global simulations, non-axisymmetry is also allowed. These simulations therefore not only allow us to study the cycle asymmetry as a function of rotation or cycle period, but also to study the effect of the degree of non-axisymmetry on it. By comparing skewness and axisymmetry of global simulations to observational data, it might be deduced whether cycles in real stars are dominated by axisymmetric or non-axisymmetric modes. Although the parameters in the simulations are still far removed from the real stellar conditions, this may provide a diagnostic tool in the future to further classify observational data into axis- and non-axisymmetric modes.

We note, however, that even the global runs have a lower average skewness than the observed MW cycles. When we assume that the observed chromospheric emission is directly proportional to the magnetic field strength, it is thus plausible that some ingredient is still missing in the simulations, which causes the asymmetry in the observed cycles. The simulations are, for example, still in a parameter regime that is too mildly turbulent, and they do not include realistic photospheres or chromospheres. The other alternative is that cycles are more symmetric for more rapidly rotating stars (for which there is a weak correlation in the MW data). In this case, the different parameter regime of the observations and simulations might explain their difference because the rotation was much faster in most of the simulated runs than in the observed stars.

\begin{figure}
   \centering
   \includegraphics[width=\hsize]{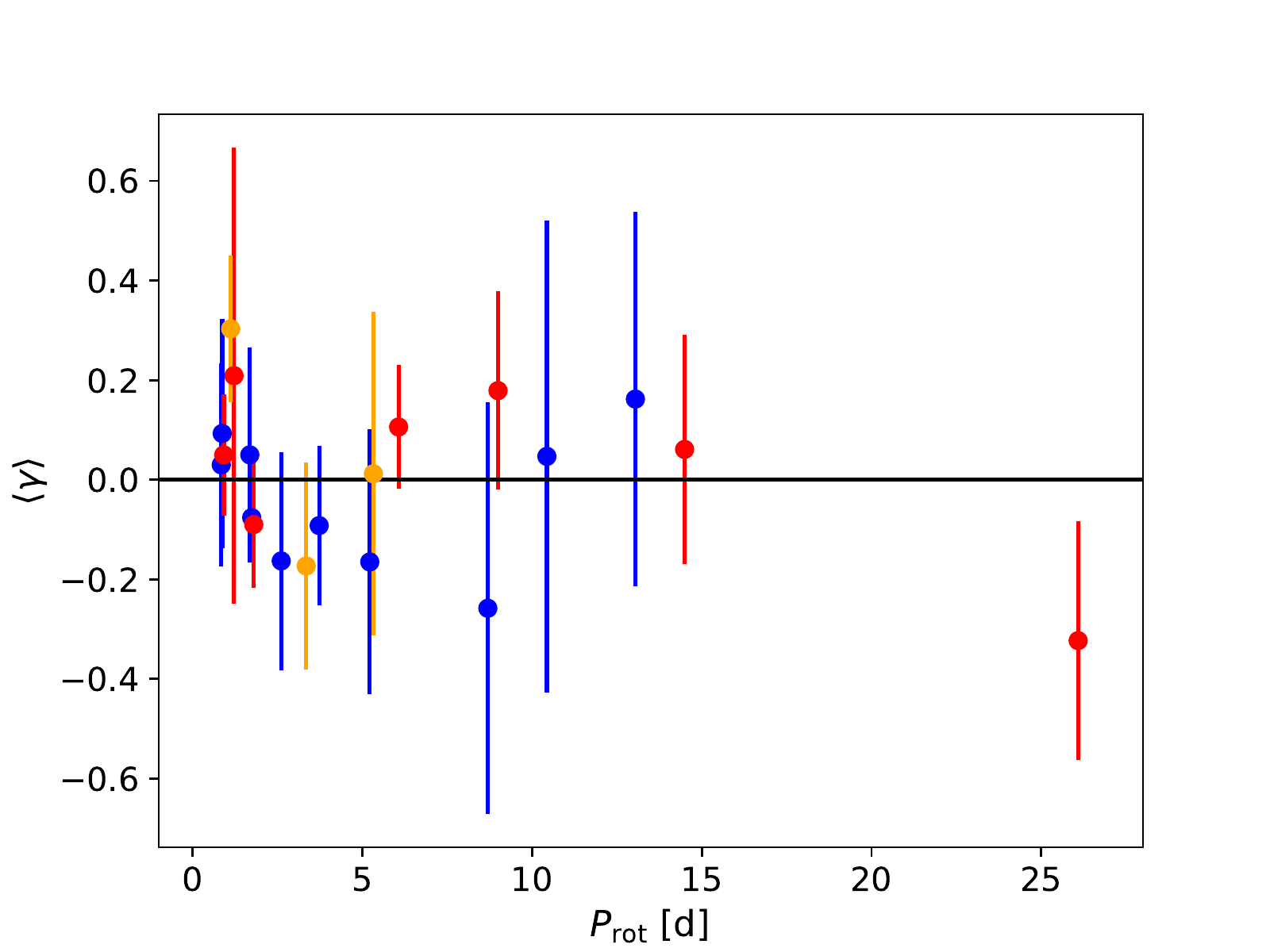}
   \caption{Mean skewness of the simulated cycles as a function of $P_{\rm{rot}}$. The error bars represent the standard deviation of the cycles in the run. Global runs are shown in red, with the high-resolution runs in orange and the wedge runs in blue. The horizontal line represents $\gamma = 0$.}
   \label{Mariangela_Prot}
\end{figure}

\begin{figure}
   \centering
   \includegraphics[width=\hsize]{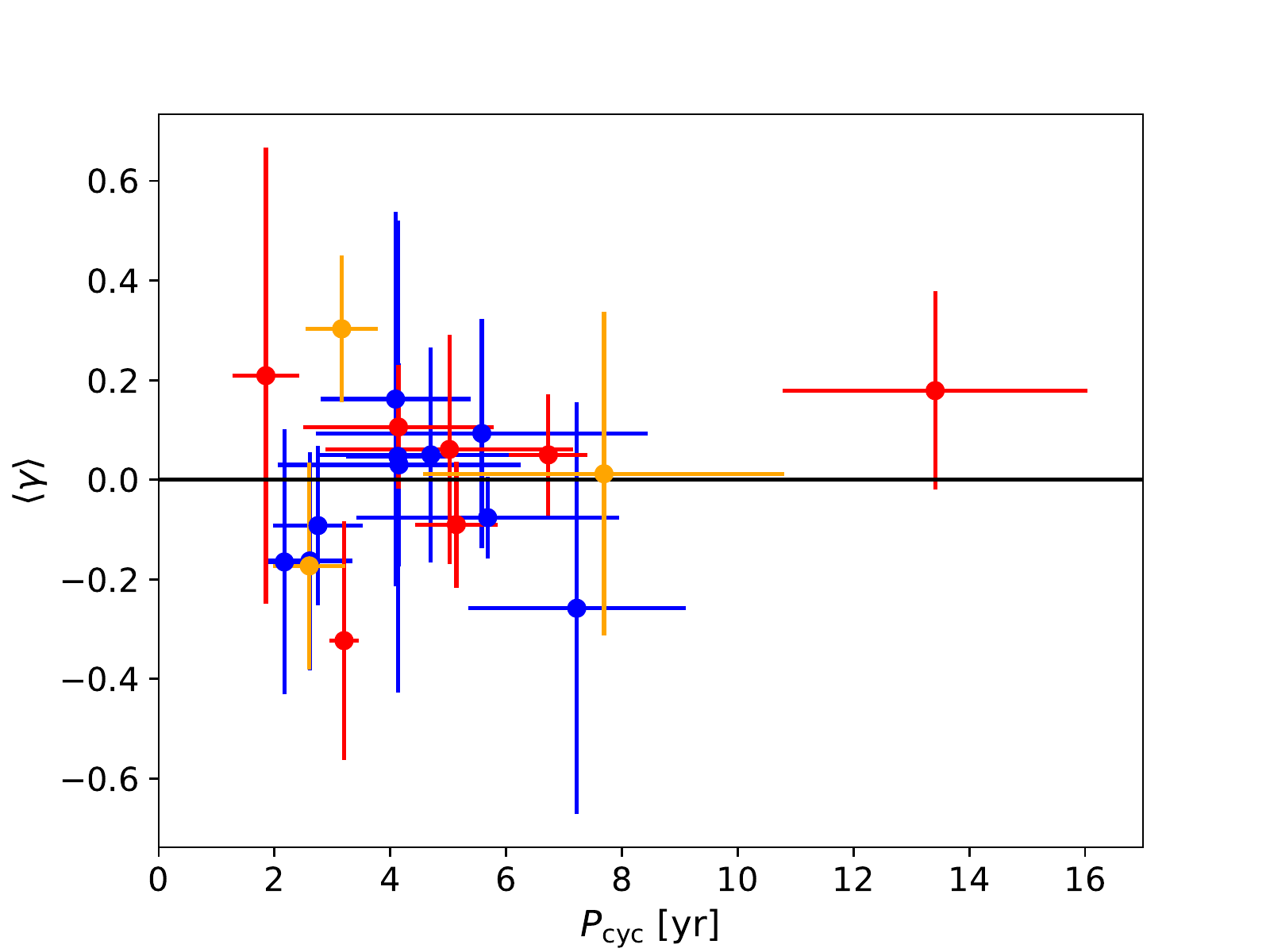}
   \caption{Same as Fig. \ref{Mariangela_Prot}, but for $P_{\rm{cyc}}$.}
   \label{Mariangela_Pcyc}
\end{figure}

Similarly to the observational data, we also compared the mean skewness of each run to other stellar parameters. Table \ref{mariangela} shows the mean skewness of all these runs, and the rotation rate of the simulated star, normalised to the solar rotation rate $\tilde{\Omega}$. The rotation rate is transformed into the rotation period by $P_{\rm{rot}} = P_{\odot}/ \tilde{\Omega}$, where $P_{\odot} = 26.09$ d is the rotational period of the Sun. $\langle \gamma \rangle$ is plotted against $P_{\rm{rot}}$ in Fig. \ref{Mariangela_Prot}, and against $P_{\rm{cyc}}$ in Fig. \ref{Mariangela_Pcyc}. Global and wedge simulations are separated by colour in the figures, as are the higher resolution global runs. We calculated Pearson correlation coefficients between $\langle \gamma \rangle$ and $P_{\rm{rot}}$, and $\langle \gamma \rangle$ and $P_{\rm{cyc}}$ for all simulations together and separately for the global and wedge runs. These are shown in Table \ref{corr_coef_sim}. The strongest correlation is $r = -0.57$ for $P_{\rm{rot}}$ for the global simulations, although this is fairly weak. Moreover, the correlation is positive for the wedge runs. For $P_{\rm{cyc}}$ the correlations are even weaker. We draw no other conclusions from this, except for the lack of strong correlations between cycle asymmetry and other parameters, as was the case with observed cycles as well.

We note that the cycle period, which we defined from the times of minima, was determined differently by \citet{viviani2018}. The authors counted how many times the mean magnetic energy level is crossed. The cycle period is also different in \citet{warnecke2018}, who determined the period using power spectra.

It must also be noted that the rotation rate, although the most relevant parameter, is not the only parameter that was varied between the simulations. Other input parameters that were changed between the runs are the grid resolution, the fluid, subgrid-scale, and magnetic Prandtl numbers, the Taylor number, and the Rayleigh number. We did not analyse, however, how these affect the cycle asymmetry because these parameters are not known for real individual stars.

\begin{table*}
\caption{Skewness of the simulated cycles. The runs are named as in the corresponding reference.}             
\label{mariangela}      
\centering                          
\begin{tabular}{c c c c c c c c c}        
\hline\hline                 
Run & $n_{\rm{cyc}}$ & $\langle \gamma \rangle$ & $\sigma$ & $\tilde{\Omega}$ & $P_{\rm{rot}}$ [d] & $P_{\rm{cyc}}$ [yr] & G/W & Reference \\    
\hline                        
  A1 & 2 & -0.323 & 0.239 & 1.0 & 26.09 & 3.20 $\pm$ 0.25 & G & \citet{viviani2018} \\
  C2 & 7 & 0.061 & 0.230 & 1.8 & 14.49 & 5.02 $\pm$ 2.14 & G & \citet{viviani2018} \\
  E & 2 & 0.179 & 0.199 & 2.9 & 9.00 & 13.41 $\pm$ 2.63 & G & \citet{viviani2018} \\
  F1 & 4 & 0.106 & 0.124 & 4.3 & 6.07 & 4.14 $\pm$ 1.65 & G & \citet{viviani2018} \\
  G$^a$ & 4 & 0.012 & 0.325 & 4.9 & 5.32 & 7.69 $\pm$ 3.12 & G & \citet{viviani2018} \\
  H$^a$ & 6 & -0.173 & 0.208 & 7.8 & 3.34 & 2.60 $\pm$ 0.62 & G & \citet{viviani2018} \\
  J & 2 & -0.090 & 0.127 & 14.5 & 1.80 & 5.14 $\pm$ 0.71 & G & \citet{viviani2018} \\
  K1 & 14 & 0.209 & 0.458 & 21.4 & 1.22 & 1.85 $\pm$ 0.58 & G & \citet{viviani2018} \\
  L$^a$ & 3 & 0.303 & 0.148 & 23.3 & 1.12 & 3.16 $\pm$ 0.62 & G & \citet{viviani2018} \\
  M & 4 & 0.050 & 0.122 & 28.5 & 0.92 & 6.73 $\pm$ 0.68 & G & \citet{viviani2018} \\
  M2 & 10 & 0.162 & 0.376 & 2.0 & 13.05 & 4.09 $\pm$ 1.30 & W & \citet{warnecke2018} \\
  M2.5 & 5 & 0.047 & 0.474 & 2.5 & 10.44 & 4.13 $\pm$ 0.90 & W & \citet{warnecke2018} \\
  M3 & 3 & -0.258 & 0.413 & 3.0 & 8.70 & 7.22 $\pm$ 1.88 & W & \citet{warnecke2018} \\
  M5 & 13 & -0.165 & 0.266 & 5.0 & 5.22 & 2.17 $\pm$ 0.35 & W & \citet{warnecke2018} \\
  M7 & 12 & -0.092 & 0.160 & 7.0 & 3.73 & 2.75 $\pm$ 0.78 & W & \citet{warnecke2018} \\
  M10 & 13 & -0.163 & 0.219 & 10.0 & 2.61 & 2.61 $\pm$ 0.73 & W & \citet{warnecke2018} \\
  M15 & 4 & -0.076 & 0.082 & 15.0 & 1.74 & 5.68 $\pm$ 2.27 & W & \citet{warnecke2018} \\
  J$^W$ & 9 & 0.050 & 0.216 & 15.5 & 1.68 & 4.70 $\pm$ 1.97 & W & \citet{viviani2018} \\
  M30 & 6 & 0.093 & 0.230 & 30.0 & 0.87 & 5.58 $\pm$ 2.87 & W & \citet{warnecke2019arxiv} \\
  M$^W$ & 10 & 0.030 & 0.204 & 31.0 & 0.84 & 4.15 $\pm$ 2.10 & W & \citet{viviani2018} \\
\hline  
\end{tabular}
\tablefoot{The G/W column divides the runs into global (G) and wedge (W) runs. The high-resolution runs are named with the superscript $a$.}
\end{table*}

\begin{table}
\caption{Correlation coefficients between $\langle \gamma \rangle$ and other parameters in the simulated cycles.}             
\label{corr_coef_sim}      
\centering                          
\begin{tabular}{c c c c c}        
\hline\hline                 
Parameter & Correlation coefficient & $p$-value \\    
\hline                        
  $P_{\rm{rot, all}}$ & -0.32 & 0.17 \\
  $P_{\rm{rot, global}}$ & -0.57 & 0.09 \\
  $P_{\rm{rot, wedge}}$ & 0.12 & 0.75 \\
  $P_{\rm{cyc, all}}$ & 0.19 & 0.11 \\
  $P_{\rm{cyc, global}}$ & 0.21 & 0.59 \\
  $P_{\rm{cyc, wedge}}$ & $3.6 \times 10^{-3}$ & 0.99 \\
\hline
\end{tabular}
\end{table}

Table \ref{comp} summarizes the main features of the simulated cycles, both including all cycles and when the global and wedge runs are separated.
Despite the low number of statistics and the different parameter space between the observations and simulations, the comparison is deemed useful. Firstly, this gives a realistic view on the current state of direct numerical simulations. Furthermore, comparing the observed and simulated trends in skewness may serve as an additional tool for deducing what type of dynamo is operating in a star.

\begin{table}
\caption{Average skewness and its standard deviation of the observed and simulated cycles.}             
\label{comp}      
\centering                          
\begin{tabular}{c c c c c}        
\hline\hline                 
Parameter & MW & All simulated runs & Global & Wedge \\    
\hline                        
  $\langle \gamma \rangle$ & 0.11 & 0.00 & 0.06 & -0.06 \\
  $\sigma$ & 0.28 & 0.32 & 0.31 & 0.31 \\
\hline
\end{tabular}
\end{table}


\section{Conclusions}

We draw the following conclusions from our study. A fast rise and slower decline is common for stellar activity cycles. The Sun has particularily asymmetric cycles. More active stars might have less asymmetric cycles, but the correlation between the skewness and other parmeters is mostly unclear. Individual cycles might have very irregular shapes, but the average cycle shape is fairly well represented with a sinusoid. The average cycle still reaches its maximum before the sinusoid because it is asymmetrical. The chromospheric and sunspot cycles do not have exactly the same shape. This means that MW cycles for other stars can probably not be directly compared to the sunspot cycle.

The numerically simulated cycles, with shorter rotation periods than the observed real stars, have on average more symmetric cycles, with a distribution in the skewness values centred very close on zero. Perhaps the simulations miss something that makes the cycles asymmetric in real stars. This might indicate that the physics that is still not captured by these models, such as the missing photosphere and chromosphere, is crucial for creating the cycle asymmetries. Other explanations for this might be a difference in the cycles between slow and fast rotators, for which there is some support from the weak correlation between the skewness and the rotation period, and the stronger anti-correlation between the skewness and $\log R'_{\rm{HK}}$ in the MW data. The simulation geometry affects the asymmetry of the simulated cycles, with the wedge simulations having on average more negatively skewed cycles than the global simulations.

\begin{acknowledgements}
  The HK\_Project\_v1995\_NSO data derive from the Mount Wilson Observatory HK Project, which was supported by both public and private funds through the Carnegie Observatories, the Mount Wilson Institute, and the Harvard-Smithsonian Center for Astrophysics starting in 1966 and continuing for over 36 years.  These data are the result of the dedicated work of O. Wilson, A. Vaughan, G. Preston, D. Duncan, S. Baliunas, and many others.
This work has made use of data from the European Space Agency (ESA) mission
{\it Gaia} (\url{https://www.cosmos.esa.int/gaia}), processed by the {\it Gaia}
Data Processing and Analysis Consortium (DPAC,
\url{https://www.cosmos.esa.int/web/gaia/dpac/consortium}). Funding for the DPAC
has been provided by national institutions, in particular the institutions
participating in the {\it Gaia} Multilateral Agreement.
TW acknowledges the financial support from the Alfred Kordelin Foundation, and thanks Angie Breimann for our discussion during the BCool meeting in Exeter.
TH acknowledges the financial support from the Academy of Finland for the project SOLSTICE (decision No. 324161).
MJK and NO acknowledge the support of the Academy of Finland ReSoLVE Centre of Excellence (grant No.~307411). This project has received funding from the European Research Council under the European Union's Horizon 2020 research and innovation programme (project "UniSDyn", grant agreement n:o 818665). MV was enrolled in the International Max Planck Research School for Solar System Science at the University of G\"ottingen.
\end{acknowledgements}

%
   \bibliographystyle{aa} 
   \bibliography{cycleshapes} 
%



\begin{appendix}
\label{append}
\section{Minima and maxima of individual MW cycles}

\begin{table*}

\caption{\label{append1} Times of minima and maxima for the MW stars and the time intervals we used to derive these as upper and lower index, and period and skewness for each cycle.} 
\centering                          
{\renewcommand{\arraystretch}{1.2} 
\begin{tabular}{c c c c c c}        
\hline\hline                 
  HD & $\#_{\rm{cyc}}$ & $t_{\rm{min}}$ & $t_{\rm{max}}$ & $P_{\rm{cyc}}$ [yr] & $\gamma$ \\    
\hline                        
  3651 & 1 & -2600$^{-1000}_{-4000}$, 2900$^{5000}_{1000}$ & -550$^{500}_{-1500}$ & 15.06 & 0.368 \\      
  4628 & 1 & -2450$^{-1500}_{-3500}$, 400$^{1300}_{-500}$ & -1000$^{0}_{-2000}$ & 7.80 & -0.049 \\
  4628 & 2 & 400, 3550$^{4700}_{2500}$ & 1800$^{2800}_{800}$ & 8.62 & 0.242 \\
  16160 & 1 & -900$_{-2000}^{500}$, 3550$_{2500}^{4500}$ & 1350$_{500}^{2300}$ & 12.18 & 0.074 \\
  26965 & 1 & -2050$_{-3200}^{-1000}$, 1700$_{1000}^{2500}$ & -300$_{-1700}^{1300}$ & 10.27 & -0.055 \\
  26965 & 2 & 1700, 5500$_{4800}^{6000}$ & 3450$_{2500}^{4500}$ & 10.40 & 0.207 \\
  32147 & 1 & -1050$_{-1800}^{-200}$, 2750$_{2000}^{3700}$ & -3150$_{-4300}^{-2200}$, 300$_{-600}^{1100}$, 4850$_{4200}^{5500}$ & 10.40 & 0.127 \\
  166620 & 1 & -2200$_{-3300}^{-800}$, 3400$_{2200}^{4500}$ & 150$_{-1500}^{1000}$ & 15.33 & 0.146 \\
  219834A & 1 & 150$_{-800}^{1000}$, 3050$_{2200}^{3600}$ & 1000$_{300}^{1800}$ & 7.94 & 0.231 \\
  219834A & 2 & 3050, 4650$_{4000}^{5500}$ & 3550$_{3000}^{4500}$ & 4.38 & 0.496 \\
  219834B & 1 & -3950$_{-4500}^{-3000}$, -500$_{-1500}^{1000}$ & -2700$_{-4000}^{-1900}$ & 9.45 & 0.521 \\
  219834B & 2 & -500, 3150$_{2600}^{3600}$ & 1100$_{300}^{1800}$ & 9.99 & 0.197 \\
  219834B & 3 & 3150, 5750$_{5000}^{6000}$ & 4200$_{3600}^{4700}$ & 7.12 & 0.238 \\
  Sun & 1 & -1150, 2650 & 200 & 10.40 & 0.338 \\
  Sun & 2 & 2650, 6200 & 3700 & 9.72 & 0.231 \\
  Sun & 3 & 6200, 10800 & 7650 & 12.59 & 0.614 \\
  10476 & 3 & -2900$_{-3600}^{-2100}$, 900$_{400}^{1400}$ & -1350$_{-2200}^{-500}$ & 10.40 & 0.109 \\
  10476 & 2 & 900, 4550$_{3700}^{5100}$ & 2100$_{1500}^{2600}$ & 9.99 & 0.446 \\
  10476 & 3 & 4550, 8300$_{7500}^{8500}$ & 6400$_{5800}^{7000}$ & 10.27 & -0.224 \\
  81809 & 1 & -2650$_{-3200}^{-2100}$, 350$_{-500}^{1000}$ & -1250$_{-2000}^{-500}$ & 8.21 & 0.128 \\
  81809 & 2 & 350, 3450$_{3000}^{4000}$ & 1600$_{900}^{2300}$ & 8.49 & 0.310 \\
  81809 & 3 & 3450, 6050$_{5200}^{6700}$ & 4500$_{4100}^{5200}$ & 7.12 & 0.228 \\
  103095 & 1 & -1700$_{-2400}^{-1000}$, 750$_{-300}^{1600}$ & -600$_{-1300}^{300}$ & 6.71 & 0.260 \\
  103095 & 2 & 750, 3500$_{3100}^{3800}$ & 2100$_{1600}^{2700}$ & 7.53 & 0.018 \\
  103095 & 3 & 3500, 5900$_{5000}^{6800}$ & 4600$_{3800}^{5300}$,7300$_{6800}^{7800}$ & 6.57 & 0.328 \\
  114710 & 1 & 1200$_{500}^{1800}$, 3350$_{2700}^{3800}$ & 1750$_{1300}^{2300}$ & 5.89 & 0.761 \\
  114710 & 2 & 3350, 5300$_{4700}^{5800}$ & 4300$_{3500}^{4900}$ & 5.34 & -0.248 \\
  114710 & 3 & 5300, 7150$_{6400}^{8100}$ & 6200$_{5700}^{6700}$ & 5.07 & -0.193 \\
  115404 & 1 & -1100$_{-2000}^{-300}$, 3500$_{3200}^{4000}$ & 400$_{-800}^{1500}$ & 12.59 & 0.130 \\
  115404 & 2 & 3500, 6650$_{6100}^{7100}$ & 5300$_{4600}^{6000}$ & 8.62 & 0.186 \\
  149661 & 1 & -3350$_{-3800}^{-2800}$, -1700$_{-2300}^{-1300}$ & -2400$_{-3000}^{-1800}$ & 4.52 & 0.006 \\
  149661 & 2 & -1700, -400$_{-1200}^{200}$ & -1300$_{-1900}^{-500}$ & 3.56 & 0.524 \\
  149661 & 3 & -400, 1300$_{800}^{1800}$ & 400$_{-300}^{1300}$ & 4.65 & -0.123 \\
  149661 & 4 & 1300, 2500$_{2100}^{3100}$ & 2000$_{1300}^{2800}$ & 3.29 & -0.114 \\
  149661 & 5 & 2500, 5100$_{4200}^{5700}$ & 3400$_{2500}^{4000}$ & 7.12 & 0.268 \\
  149661 & 6 & 5100, 6900$_{6500}^{7400}$ & 6250$_{5800}^{6800}$ & 4.93 & -0.638 \\
  152391 & 1 & -1800$_{-3200}^{-500}$, 2100$_{1600}^{2800}$ & 200$_{-500}^{800}$ & 10.68 & -0.460 \\
  152391 & 2 & 2100, 5650$_{5000}^{6100}$ & 4050$_{3200}^{4900}$ & 9.72 & -0.038 \\
  152391 & 3 & 5650, 7700$_{7200}^{8200}$ & 6350$_{5800}^{6800}$ & 5.61 & 0.379 \\
  160346 & 1 & -3750$_{-4100}^{-3000}$, -1100$_{-1600}^{-500}$ & -2700$_{-3300}^{-2000}$ & 7.26 & 0.008 \\
  160346 & 2 & -1100, 1400$_{1000}^{2000}$ & 150$_{-500}^{700}$ & 6.84 & 0.178 \\
  160346 & 3 & 1400, 3950$_{3200}^{4600}$ & 2350$_{1800}^{2800}$ & 6.98 & 0.077 \\
  160346 & 4 & 3950, 6700$_{6200}^{7200}$ & 5250$_{4700}^{5700}$ & 7.53 & 0.175 \\
  201091 & 1 & -3400$_{-3800}^{-3100}$, -850$_{-1400}^{-400}$ & -2300$_{-2800}^{-1800}$ & 6.98 & -0.085 \\
  201091 & 2 & -850, 2150$_{1800}^{2800}$ & 400$_{0}^{1000}$ & 8.21 & -0.004 \\
  201091 & 3 & 2150, 4500$_{4000}^{5100}$ & 3400$_{2500}^{4000}$ & 6.43 & 0.071 \\
  201091 & 4 & 4500, 6900$_{6400}^{7600}$ & 5700$_{5000}^{6500}$ & 6.57 & -0.113 \\
\hline                                   
\end{tabular}
\tablefoot{[$t_{\rm{min/max}}$]=JD-2444000. $t_{\rm{min}}$ and $t_{\rm{max}}$ for the Sun are from \citet{hathaway2015}. Intervals used in the fitting of minima are only listed once for each minimum. With HD 32147 and HD 103095, the additional maxima were used in to increase statistics in the calculation of $\langle t_{\rm{r}} \rangle / \langle t_{\rm{d}} \rangle$.}
}
\end{table*}

\end{appendix}

\end{document}